\documentclass[useAMS,usenatbib]{mn2e}
\usepackage{graphicx}
\usepackage{color}
\usepackage{subfigure}
 \newcommand{\Msolar}{\mbox{\,$\rm M_{\odot}$}}        % solar mass
 \newcommand{\Lsolar}{\mbox{\,$\rm L_{\odot}$}}        % solar luminosity
 \newcommand{\Teff}{\mbox{\,\em $T_{\rm eff}$}}      % effective temperature
 \newcommand{\logg}{\mbox{\,log $g$}}                   % surface gravity
 \newcommand{\kelvin}{\,\mbox{K}}                       % K Kelvin

\title[HD\,179821 (V1427\,Aql, IRAS\,19114+0002) -- A Massive Post-Red Supergiant Star?]{HD\,179821 (V1427\,Aql, IRAS\,19114+0002) -- A 
Massive Post-Red Supergiant Star?}
\author[T. \c{S}ahin, David L. Lambert, Valentina G. Klochkova and Vladimir E. Panchuk]{T.
\c{S}ahin$^{1,2}$\thanks{E-mail:timursahin@akdeniz.edu.tr(TS); dll@astro.as.utexas.edu(DLL);
valenta@sao.ru(VGK)}, David L. Lambert$^{2}$, Valentina G. Klochkova$^{3}$, and Vladimir E. Panchuk$^{3, 4}$ \\\\
$^{1}$ Akdeniz University, Faculty of Science, Space Sciences and Technologies
Department 07058, Antalya, Turkey\\
$^{2}$ Department of Astronomy and The W. J. McDonald Observatory, The University of Texas, Austin, Texas, USA
78712 \\
$^{3}$ Special Astrophysical Observatory, Nizhnij Arkhyz, Karachai-Cherkessia, 369167 Russia\\
$^{4}$ National Research  University ITMO, St.-Petersburg, 197101, Russia 
}

\begin{document}

\pagerange{\pageref{firstpage}--\pageref{lastpage}} \pubyear{2013}

\maketitle

\label{firstpage}

\begin{abstract}

We have derived elemental abundances of a remarkable star, HD\,179821, with unusual composition (e.g.
[Na/Fe]=1.0$\pm$0.2 dex) and extra-ordinary spectral characteristics. Its metallicity at [Fe/H]=0.4 dex places
it among the most metal-rich stars yet analyzed. The abundance analysis of this luminous
star is based on high resolution and high quality (S/N$\approx$120--420) optical echelle spectra from
McDonald Observatory and Special Astronomy Observatory. The
data includes five years of observations over twenty-one epochs. Standard 1D {\sc LTE}
analysis provides a fresh determination of the atmospheric parameters over all epochs: \Teff =
7350$\pm$200 \kelvin, \logg = +0.6$\pm$0.3, and a microturbulent velocity $\xi =$ 6.6$\pm$1.6 km
s$^{\rm -1}$ and [Fe/H] = 0.4$\pm$0.2, and a carbon abundance [C/Fe]= $-$0.19$\pm$0.30. We find oxygen
abundance [O/Fe]= $-$0.25$\pm$0.28 and an enhancement of 0.9 dex in N. A supersonic macroturbulent velocity of
22.0 $\pm$ 2.0 km s$^{\rm -1}$ is determined from both strong and weak Fe\,{\sc i} and Fe\,{\sc ii}
lines. Elemental abundances are obtained for 22 elements. HD 179821 is not enriched in s-process
products. Eu is overabundant relative to the anticipated [X/Fe] $\approx$ 0.0. Some peculiarities of
its optical spectrum (e.g. variability in the spectral line shapes) is noticed. This includes the line
profile variations for H$\alpha$ line. Based on its estimated luminosity, effective temperature and surface gravity,
HD\,179821 is a massive star evolving to become a red supergiant and finally a Type II
supernova.      

\end{abstract}


\section{INTRODUCTION}

The spectral class of F-G luminous giants may encompass stars on two different
evolutionary paths. Some  stars may be massive stars evolving from the main sequence and
some of these massive stars may now be in a post-red supergiant phase. Alternatively,
other stars may be departing the asymptotic giant branch (AGB) evolving at roughly
constant luminosity to hotter temperatures and the tip of the white dwarf cooling track.
Unambiguous assignment of a F-G supergiant to the proper evolutionary path is not always
immediately possible, even when a wide variety of observational techniques are applied
and the electromagnetic spectrum is well sampled.

HD\,179821, also known as V1427 Aql and IRAS\,19114+0002, remains a supergiant of
uncertain heritage despite a lengthy literature and frequent investigations into its
status. Advocates for a post-AGB origin include Za\v{c}s et al. (1996) and Reddy \&
Hrivnak (1999) who gave weight to their measurements of  overabundances of $s$-process
nuclides. Others have stressed the star's distance as implied by its radial velocity and
characteristics of its circumstellar gas and dust shell in suggesting that the star is a
massive supergiant: see, for example, Jura \& Werner (1999) and Jura, Velusamy, \&
Werner (2001) with the latter paper carrying the provocative title `What next for the
likely presupernova HD 179821?'. Oudmaijer et al. (2009) confidently place HD\,179821
among massive stars in a post-red supergiant phase.

An oft stated assertion is that a star's chemical composition provides clues to its evolutionary
history. Certainly, one anticipates readily observable distinctive signatures between an evolved
massive star and a mature post-AGB star (i.e., a star that left the AGB after experiencing many
thermal pulses and extensive third dredge-up episodes). The massive star will have experienced
mixing between envelope and interior at a minimum and, thus, readjustment of its surface C, N
and possibly O abundances: a decrease in C abundance with an offsetting increase in N abundance
is assured but with very few exceptions (Na, possibly) all other elements will retain their
natal abundances. On the other hand, a mature post-AGB star will be markedly $s$-process 
enriched with a likely enrichment of C accompanying the $s$-process enrichment. The contrasting
compositions of massive and post-AGB stars surely represent a testable proposition.
Unfortunately, some stars appear to evolve off the AGB before the third dredge-up has altered
the surface abundances of the $s$-process nuclides (see De Ruyter et al.
2006, and references therein).

The paper is organized as follows: Section 2 discusses the high-resolution optical
spectra obtained at the two observatories from 2008--2013; General properties of the
spectra are discussed briefly in Section 3; Section 4 describes our abundance analysis
and reanalyses previous analyses; Section 5 in a collection of concluding remarks places
HD\,179821 in its likely evolutionary status as a massive star evolving to become a red
supergiant and finally a Type II supernova.     

\section{Observations}

Our investigation of HD 179821 is based on high-resolution optical
spectra obtained between 2008 and 2013 at two observatories: the McDonald Observatory (McD)
and the Special Astrophysical Observatory (SAO). 
The log of observations is given in Table 1.

The character of {\sc HD 179821}'s spectrum is displayed by Figure~1. Some
features are blended and require spectrum synthesis to yield useful abundance
information. Many lines suitable for abundance analysis are apparently
unblended.

\begin{table}
\begin{center}
\caption{Log of observations for HD\,179821 obtained at the McDonald and the SAO. S/N ratios for the {\sc SAO}
spectra presents the S/N values at 5500 \AA\, while in the {\sc McDonald} spectra, the S/N values in the raw spectra are reported
near 5000 \AA.}
$$
\begin{array}{l|@{}c@{}|l|c|c}
\hline
$Obs. Period$   &$Exposure$ &$Wave.Range$         &$S/N$  &$Notes$\\
\cline{2-3}
              &  $(sec)$  & \hskip 0.8cm $\AA$ &     &     \\
\hline \hline
21\,Apr\,2008 & 1 x 1800& 3877-10\,338      & 174 & McD \\
13\,Jun\,2008 & 1 x 1200& 3832-10\,339      & 193 & McD \\
11\,Jul\,2008 & 1 x 1800& 3833-10\,337      & 136 & McD \\
10\,Aug\,2008 & 1 x 1200& 3833-10\,337      & 154 & McD \\
17\,Aug\,2008 & 3 x 2400& 4550-6000\,\,\,\, & 252 & SAO \\
17\,Sep\,2008 & 3 x 1800& 5260-6766\,\,\,\, & 420 & SAO \\
13\,Apr\,2009 & 4 x 2400& 5214-6688\,\,\,\, & 385 & SAO \\
09\,May\,2009 & 1 x 1200& 3832-10\,338      & 172 & McD \\
09\,May\,2009 & 1 x 1200& 3832-10\,338      & 172 & McD \\
07\,Nov\,2009 & 3 x 3000& 5220-6690\,\,\,\, & 119 & SAO \\
21\,Nov\,2009 & 1 x 1500& 3832-10\,337      & 166 & McD \\ 
22\,May\,2010 & 1 x 1800& 3833-10\,337      & 123 & McD \\
22\,May\,2010 & 1 x 1800& 3833-10\,337      & 123 & McD \\
03\,Jun\,2010 & 3 x 2400& 5210-6690\,\,\,\, & 210 & SAO \\
30\,Jul\,2010 & 3 x 2700& 4422-5930\,\,\,\, & 324 & SAO \\
24\,Sep\,2010 & 3 x 1800& 5220-6690\,\,\,\, & 285 & SAO \\
17\,Nov\,2010 & 3 x 2400& 5160-6689\,\,\,\, & 352 & SAO \\
16\,May\,2011 & 1 x 1500& 3832-10\,338      & 208 & McD \\ 
30\,May\,2013 & 4 x 0900& 3916-6980         & 260 & SAO \\      
27\,Aug\,2013 & 4 x 2400& 3917-6980         & 283 & SAO \\	
08\,Oct\,2013 & 3 x 2400& 3914-6077         & 201 & SAO \\
\hline \hline
\end{array}
$$
\end{center} 
\end{table}

\begin{figure*}
%\begin{center}
\includegraphics[width=179mm,height=87mm,angle=0]{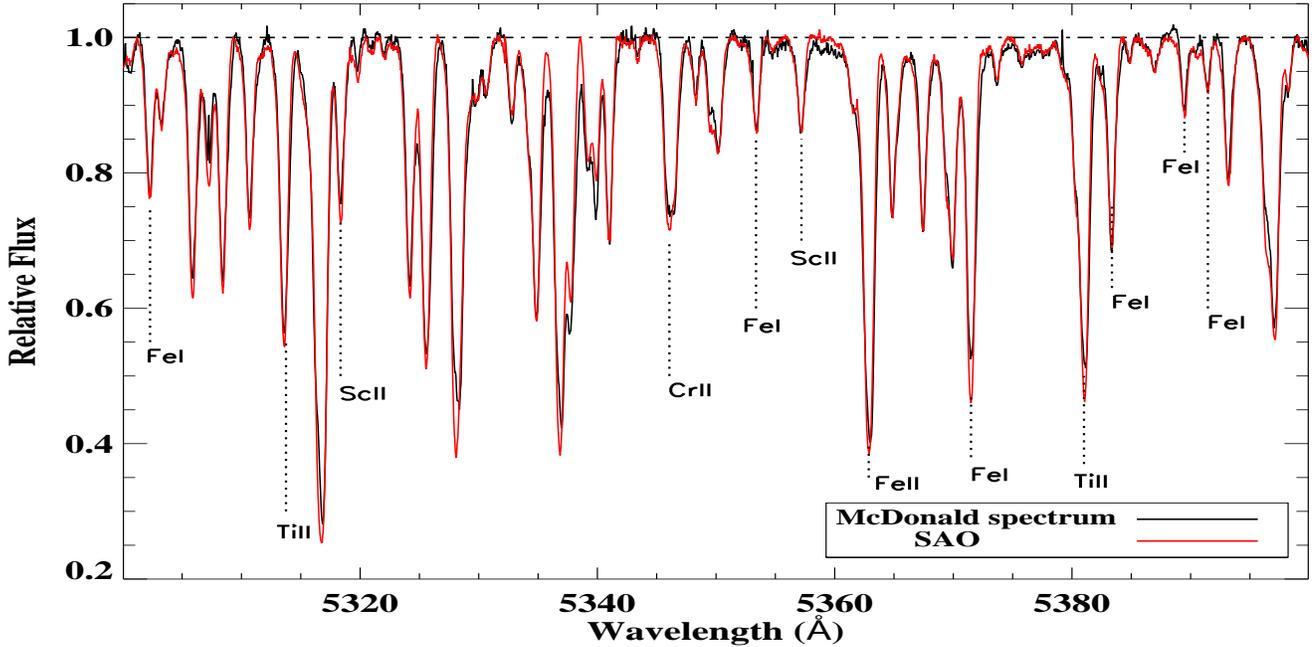}
\caption{The spectrum for HD\,179821 over the wavelength region 5300-5400 \AA\,. The McDonald spectrum (13 Jun 
2008) is plotted in black and the SAO spectrum (17 Nov 2010)
in red.}
%\end{center}
\label{spec_plot}
\end{figure*}

\subsection{McDonald Observatory's Tull spectrograph}

Spectra were obtained at the W.J. McDonald Observatory with the 2.7m Harlan J. Smith reflector
and the Tull coud\'{e} echelle spectrograph (Tull et al. 1995). A single exposure covers the
wavelength range from about 3800\AA\, to 10\,500\AA. Spectral coverage is incomplete from about
5800\AA\, to longer wavelengths. A ThAr hollow cathode lamp provided the wavelength
calibration. Exposures of 20 to 30 minutes provided a satisfactory S/N ratio over much of the
captured spectrum. 

Observations were reduced using the software package in {\sc IRAF}.\footnote { {\sc
IRAF} is distributed by the National Optical Astronomy Observatories, which is operated
by the Association of Universities for Research in Astronomy (AURA), Inc., under
cooperative agreement with the National Science Foundation.} The bias level in the over
scan area was modeled with a polynomial and subtracted. The scattered light was modeled
and removed from the spectrum. In order to correct for pixel-to-pixel sensitivity
variations, flat field exposures from a halogen lamp were used. Individual orders were
cosmic-ray cleaned, and wavelength calibrated. The internal accuracy of the wavelength
calibration via the ThAr lamp spectra was always better than on 0.003 \AA\,rms.
Rectification and merging of the individual orders into one spectrum were performed with
bespoke echelle reduction software in {\sc IDL} (\c{S}ahin 2008). The line equivalent widths
(EWs) were measured in the same manner as in \c{S}ahin et al. (2011), however, one
additional test was performed. The EWs were compared to those measured by using {\sc
STARLINK} spectrum analysis program {\sc DIPSO} (Howarth et al. 1998), {\sc
SPECTRE}\footnote{http://www.as.utexas.edu/~chris/spectre.html} (Sneden 1973), and an
in-house developed {\sc IDL} package to check any systematic errors due to continuum
placement in those measurements. The results for a representative sample of weak and
strong lines agreed well within $\pm$5 m\AA. In equivalent width measurements, the local
continuum was fitted with a first-degree polynomial then equivalent widths were measured
using a Gaussian profile. For strong lines, a direct integration was preferred to the
Gaussian approximation. The errors for each equivalent width measurement were determined
by using the prescriptions given by Howarth \& Phillips (1986).

\begin{table}
\begin{center}
\caption{The measured heliocentric radial velocities V$_{\rm HEL}$ in the McDonald and the SAO
spectra of HD\,179821. Also, a summary of the heliocentric radial velocities of identified DIBs (SAO velocities are
corrected for an offset of 1.2 km s$^{\rm -1}$) is
presented (fourth column). The number of identified DIB features is included in parenthesis.}
$$
\begin{array}{l|@{}c@{}|c|@{}l@{}|c}
\hline
$Obs. Period$ & $HJD$ & V_{\rm HEL}  & \hskip 0.5cm $DIB$ & $Notes$   \\
\cline{2-4}
               &2450000+& \hskip 0.2cm km\,s^{\rm -1} & \hskip 0.5cm km\,s^{\rm -1} &  \\
\hline \hline
21\,Apr\,2008 & 4577.9&91.5\pm0.2  &-8.4\pm0.5   (5)  &McD \\
13\,Jun\,2008 & 4630.9&82.2\pm0.2  &-8.2\pm0.5   (5)  &McD \\
11\,Jul\,2008 & 4658.7&85.7\pm0.2  &-8.7\pm0.2   (6)  &McD \\
10\,Aug\,2008 & 4688.6&87.9\pm0.2  &-8.4\pm0.7   (5)  &McD \\
17\,Aug\,2008 & 4696.4&88.1\pm0.1  &-7.8\pm0.5  (6)   &SAO \\
17\,Sep\,2008 & 4727.3&80.0\pm0.1  &-9.4\pm0.4  (14)  &SAO \\
13\,Apr\,2009 & 4935.5&88.5\pm0.1  &-8.7\pm0.5  (9)   &SAO \\
09\,May\,2009 & 4960.9&87.2\pm0.2  &-8.3\pm0.6   (5)  &McD \\
09\,May\,2009 & 4960.9&87.3\pm0.2  &-8.7\pm0.4   (5)  &McD \\
07\,Nov\,2009 & 5143.2&84.8\pm0.2  &-8.0\pm0.4   (16) &SAO \\
21\,Nov\,2009 & 5156.5&80.5\pm0.3  &-8.5\pm0.6   (5)  &McD \\
22\,May\,2010 & 5338.9&85.9\pm0.2  &-8.3\pm0.7   (4)  &McD \\
22\,May\,2010 & 5338.9&86.7\pm0.2  &-8.3\pm0.8   (5)  &McD \\
03\,Jun\,2010 & 5351.5&82.1\pm0.1  &-9.0\pm0.3  (15)  &SAO \\
30\,Jul\,2010 & 5408.4&85.7\pm0.1  &-9.3\pm0.0  (2)   &SAO \\
24\,Sep\,2010 & 5464.2&79.9\pm0.1  &-9.0\pm0.4  (9)   &SAO \\
17\,Nov\,2010 & 5518.2&89.6\pm0.1  &-8.9\pm0.8  (9)   &SAO \\
16\,May\,2011 & 5697.9&80.7\pm0.3  &-9.1\pm0.7   (4)  &McD \\
30\,May\,2013 & 6442.5&98.1\pm0.1  &-8.3\pm0.4   (17) &SAO \\
27\,Aug\,2013 & 6532.4&86.5\pm0.1  &-8.1\pm0.5   (10) &SAO \\
08\,Oct\,2013 & 6574.3&83.5\pm0.1  &-7.7\pm0.5   (12) &SAO \\
\hline \hline
\end{array}
$$
\end{center}
\end{table}

\subsection{Special Astrophysical Observatory's Nasymth spectrograph}

Spectra were obtained with the {\sc NES} echelle spectrograph (Panchuk et al. 2007,
2009) mounted at the Nasmyth focus of the 6-m telescope of the Special Astrophysical
Observatory of the Russian Academy of Sciences (SAO RAS). Observations were made with a
2048\,$\times$\,2048 CCD and an image slicer. For the the SAO spectra in 2013, a larger CCD
(2048\,$\times$\,4096) was employed. This provided an increase in the recorded spectral range.
The spectroscopic resolution and the signal-to-noise ratio are R$\ge$60000 and S/N$\ge$100,
respectively. A modified (Yushkin \& Klochkova 2005) ECHELLE context of the MIDAS package was
used to extract one-dimensional vectors from the two-dimensional echelle spectra. Cosmic-ray
hits were removed via median averaging of two successively taken spectra. Wavelength
calibration was performed using the spectra of a hollow-cathode ThAr lamp. The NES spectra cover the wavelength range 3916\AA\ to 6980\AA\ without gaps and, thus, provide
lines that fall in the inter-order gaps in the McDonald spectra up to almost 7000\AA.

\section{General features of HD\,179821'\lowercase{s} spectrum}

A partial glimpse into the spectrum of HD 179821 is provided by Figure~1. Comparison of McDonald
and SAO spectra shows that they are of matching quality. Comparison of
equivalent widths of unblended lines from the McDonald spectrum for 2008 August 10 and the SAO
spectrum of 2008 August 17 (a one week interval in which, we assume, the star's spectrum varied
very little) shows close agreement (EW(SAO)=1.01(0.01)EW(McD)-5.95(1.94)) and, thus, spectra from
the two observatories may be combined to study the star's long-term behavior.

\subsection{H$\alpha$ profiles}

Photometric variability implies spectroscopic variability. Among the
striking indicators of the variability is the changing profiles of
H$\alpha$, a variation already noted by Kipper (2008). Figure~2
shows a selection of H$\alpha$ profiles. Weak emission in the blue or
red wings is seen on occasions but a more striking  change occurs in the
width and radial velocity of the strong absorption line, for example,
the 2010 May and July profiles are shifted to the blue by
about 20 km s$^{-1}$.

\begin{figure}
%\begin{center}
\includegraphics[width=85mm,height=70mm,angle=0]{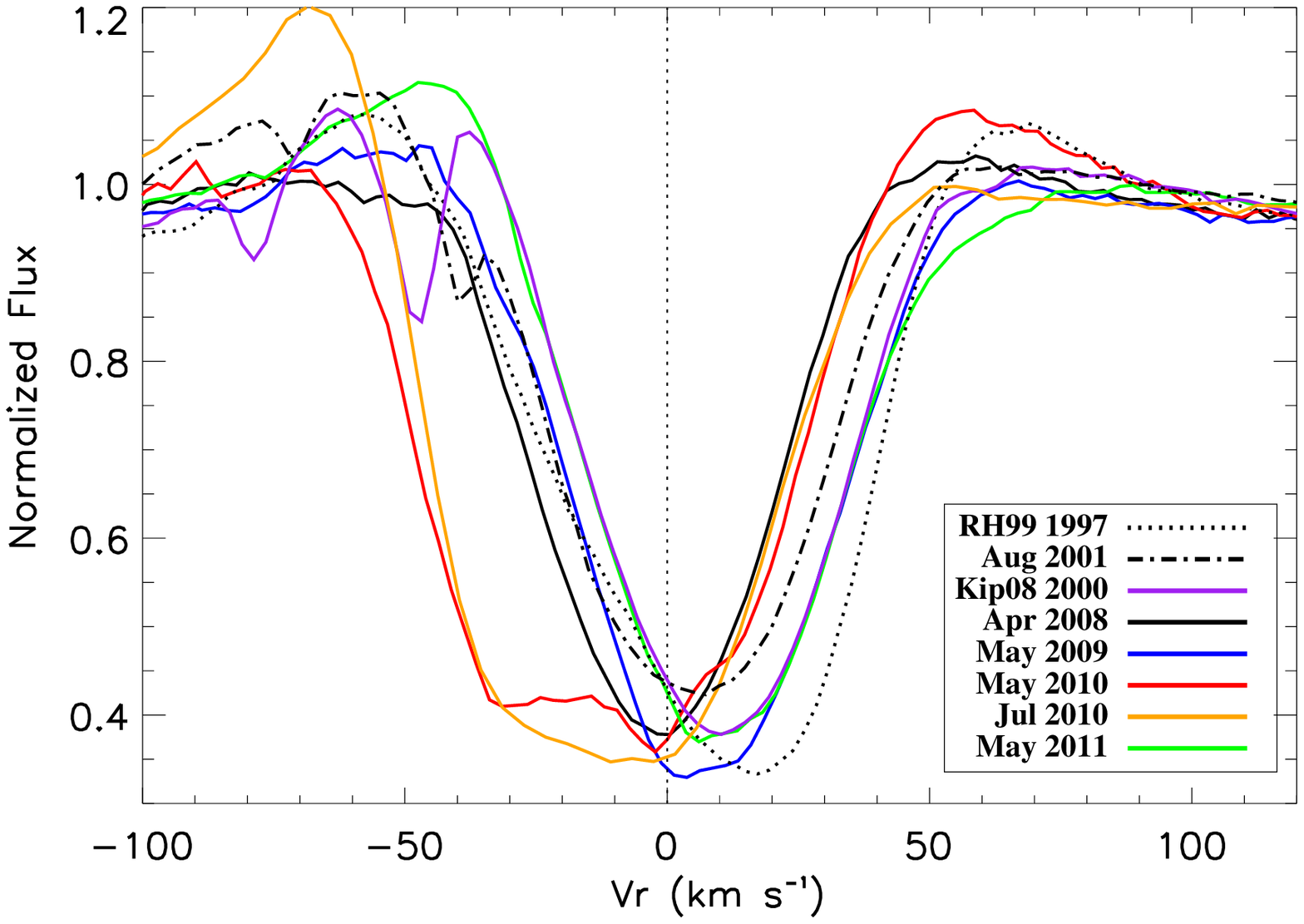}
\caption{The profile of the H$\alpha$ line obtained in several observing runs. The profiles are 
plotted relative to photospheric velocity of the individual spectra. The ELODIE and the McDonald
spectra analyzed by Kipper (2008) and Reddy \& Hrivnak (1999), respectively, are also presented. The profile of
H$\alpha$ in orange color is 21 July 2010 spectrum obtained by the Hobby Eberly Telescope (Luck
2016, private communication). The dot-dashed line shows the 6 August 2001 McDonald spectrum (Reddy
2016, private communication).}
%\end{center}
\label{h_alpha}
\end{figure}

\begin{table*}
\caption{Lines used in the analysis of McDonald and SAO spectra. Abundances for individual lines are
those obtained for the 2009 May 9 spectrum and a model of $T_{\rm eff}=7350$ K, $\log$ g = 0.64
and $\xi$=6.6 km s$^{\rm -1}$.}

$$
        \begin{array}{@{}l|@{}rrrrc@{}c||l|@{}rrrrc@{}r}
\hline           \hline
 $Species$&\lambda&$LEP$ &\log(gf)&  $EW$  &\log\epsilon(X)\,& $REF$^{\rm a}&$Species$&\lambda &$LEP$&\log(gf)& $EW$&\log\epsilon(X)\,&$REF$^{\rm a}\\
\cline{2-6}
\cline{9-13}
          &($\AA$)&($eV$)&        & (m$\AA$)&(dex)           &              &         &($\AA$) & ($eV$) &     & (m$\AA$)&($dex$)&   \\
           \hline
\hline
C\,{\sc I}  &  4775.88 &  7.49 & $-$2.304 &  115.0  &  9.14 &$WFD$   &Sc\,{\sc II}&  4431.33 &  0.61 & $-$1.970 &syn	&  4.05 &$NIST$ \\
C\,{\sc I}  &  6014.84 &  8.64 & $-$1.584 &   36.9  &  8.65 &$WFD$   &Sc\,{\sc II}&  5357.19 &  1.51 & $-$2.110 &75.0	&  3.83 &$NIST$ \\
C\,{\sc I}  &  6828.19 &  8.53 & $-$1.461 &  118.7  &  9.13 &$WFD$   &Sc\,{\sc II}&  6604.60 &  1.36 & $-$1.309 &227.3  &  3.68 &$NIST$ \\
C\,{\sc I}  &  7108.92 &  8.64 & $-$1.594 &   38.3  &  8.70 &$WFD$   &Ti\,{\sc II}&  4409.52 &  1.23 & $-$2.530 &syn	&  5.84 &$L013$ \\
C\,{\sc I}  &  7111.45 &  8.64 & $-$1.085 &   88.9  &  8.66 &$WFD$   &Ti\,{\sc II}&  4411.94 &  1.23 & $-$2.620 &syn	&  5.94 &$L013$ \\
C\,{\sc I}  &  7113.17 &  8.64 & $-$0.773 &  141.6  &  8.67 &$WFD$   &Ti\,{\sc II}&  4432.09 &  1.24 & $-$3.080 &syn	&  5.84 &$L013$ \\
C\,{\sc I}  &  7115.17 &  8.64 & $-$0.934 &  136.1  &  8.80 &$WFD$   &Ti\,{\sc II}&  4493.54 &  1.08 & $-$2.780 &syn	&  5.97 &$L013$ \\
C\,{\sc I}  &  7119.70 &  8.64 & $-$1.159 &   79.5  &  8.65 &$WFD$   &Ti\,{\sc II}&  4495.46 &  1.18 & $-$3.544 &syn	&  5.60 &$L013$ \\
C\,{\sc I}  &  7476.15 &  8.77 & $-$1.574 &   36.7  &  8.77 &$WFD$   &Ti\,{\sc II}&  5262.14 &  1.58 & $-$2.106 &syn	&  5.69 &$L013$ \\
C\,{\sc I}  &  7483.41 &  8.77 & $-$1.372 &   49.5  &  8.72 &$WFD$   &Ti\,{\sc II}&  5268.62 &  2.60 & $-$1.610 &syn	&  5.69 &$L013$ \\
N\,{\sc I}  &  7442.29 & 10.33 & $-$0.384 &   syn   &  9.95 &$WFD$   &Ti\,{\sc II}&  5418.75 &  1.58 & $-$2.110 &syn	&  5.84 &$L013$ \\
N\,{\sc I}  &  8629.23 & 10.69 &    0.075 &   syn   &  9.40 &$WFD$   &Cr\,{\sc I} &  4646.17 &  1.03 & $-$0.740 &104.7  &  6.09 &$SOLS$ \\
N\,{\sc I}  &  8594.01 & 10.68 & $-$0.335 &   syn   &  9.80 &$WFD$   &Cr\,{\sc I} &  5348.33 &  1.00 & $-$1.210 &43.9	&  6.01 &$SOLS$ \\
O\,{\sc I}  &  6156.77 & 10.74 & $-$0.694 &   syn   &  9.00 &$WFD$   &Cr\,{\sc I} &  5409.78 &  1.03 & $-$0.720 &syn	&  6.02 &$NIST$ \\
O\,{\sc I}  &  6158.18 & 10.74 & $-$0.409 &   syn   &  9.08 &$WFD$   &Cr\,{\sc II}&  5420.92 &  3.76 & $-$2.360 &syn	&  6.12 &$NIST$ \\
Na\,{\sc I} &  5682.63 &  2.10 & $-$0.706 &  syn    &  7.68 &$NIST$  &Cr\,{\sc II}&  5246.78 &  3.71 & $-$2.460 &211.9  &  6.20 &$NIST$ \\
Na\,{\sc I} &  5688.21 &  2.10 & $-$0.452 &  syn    &  7.66 &$NIST$  &Cr\,{\sc II}&  5310.70 &  4.07 & $-$2.270 &180.8  &  6.12 &$NIST$ \\
Na\,{\sc I} &  5688.19 &  2.10 & $-$1.406 &  syn    &  7.66 &$NIST$  &Cr\,{\sc II}&  5508.63 &  4.16 & $-$2.120 &209.3  &  6.19 &$NIST$ \\
Mg\,{\sc I} &  4057.48 &  4.35 & $-$1.190 &  179.8  &  8.26 &$THEV$  &Mn\,{\sc I} &  4754.04 &  2.28 & $-$0.085 &105.7  &  5.90 &$NIST$ \\
Mg\,{\sc I} &  4167.23 &  4.35 & $-$1.040 &  229.2  &  8.42 &$THEV$  &Mn\,{\sc I} &  4783.42 &  2.29 &    0.042 &122.8  &  5.87 &$NIST$ \\
Mg\,{\sc I} &  4702.99 &  4.35 & $-$0.550 &  301.0  &  8.34 &$THEV$  &Mn\,{\sc I} &  6016.64 &  3.07 & $-$0.183 &syn	&  5.83 &$NIST$ \\
Mg\,{\sc I} &  5528.41 &  4.35 & $-$0.470 &  308.6  &  8.24 &$THEV$  &Mn\,{\sc I} &  6021.80 &  3.07 &    0.035 &syn	&  5.90 &$NIST$ \\
Mg\,{\sc II}&  7877.06 &  9.99 &    0.390 &  310.0  &  8.36 &$KELP$  &Ni\,{\sc I} &  4331.64 &  1.68 & $-$2.100 &syn	&  7.15 &$NIST$ \\
Mg\,{\sc II}&  4481.13 &  8.86 &    0.749 &  syn    &  8.33 &$KELP$  &Ni\,{\sc I} &  4829.03 &  3.54 & $-$0.330 &79.6	&  6.95 &$NIST$ \\
Mg\,{\sc II}&  4481.15 &  8.86 & $-$0.553 &  syn    &  8.33 &$KELP$  &Ni\,{\sc I} &  4831.18 &  3.61 & $-$0.410 &76.3	&  7.06 &$NIST$ \\
Mg\,{\sc II}&  4481.33 &  8.86 &    0.594 &  syn    &  8.33 &$KELP$  &Ni\,{\sc I} &  5176.56 &  3.90 & $-$0.440 &53.6	&  7.10 &$NIST$ \\
Mg\,{\sc II}&10\,092.09& 11.63 &    0.910 &  syn    &  8.52 &$SCOT$  &Ni\,{\sc I} &  6176.82 &  4.09 & $-$0.529 &36.3	&  7.13 &$NIST$ \\
Al\,{\sc I} &  8772.87 &  4.02 & $-$0.349 &  syn    &  6.77 &$KELP$  &Ni\,{\sc I} &  6767.78 &  1.83 & $-$2.167 &41.8	&  7.04 &$NIST$ \\
Al\,{\sc I} &  8773.90 &  4.02 & $-$0.192 &  syn    &  6.77 &$KELP$  &Ni\,{\sc I} &  7122.21 &  3.54 & $-$0.050 &96.2	&  6.75 &$THEV$ \\
Al\,{\sc I} &  8773.91 &  4.02 & $-$1.495 &  syn    &  6.77 &$KELP$  &Ni\,{\sc II}&  4362.09 &  4.03 & $-$2.723 &syn	&  7.15 &$NIST$ \\
Si\,{\sc I} &  5645.62 &  4.93 & $-$2.141 &   35.6  &  8.61 &$KELP$  &Zn\,{\sc I} &  4722.16 &  4.03 & $-$0.390 &51.3	&  4.97 &$BIEG$ \\
Si\,{\sc I} &  5701.10 &  4.93 & $-$2.050 &   36.0  &  8.53 &$KELP$  &Zn\,{\sc I} &  4810.54 &  4.08 & $-$0.170 &103.1  &  5.18 &$BIEG$ \\
Si\,{\sc I} &  5708.40 &  4.95 & $-$1.470 &   93.3  &  8.48 &$NIST$  &Y\,{\sc II} &  5119.12 &  0.99 & $-$1.360 &96.3	&  2.78 &$HANN$ \\
Si\,{\sc I} &  5772.15 &  5.08 & $-$1.750 &   40.9  &  8.40 &$KELP$  &Y\,{\sc II} &  5200.41 &  0.99 & $-$0.570 &233.1  &  2.72 &$HANN$ \\
Si\,{\sc I} &  5797.86 &  4.95 & $-$2.051 &   64.0  &  8.84 &$KELP$  &Y\,{\sc II} &  5728.88 &  1.84 & $-$1.120 &62.0	&  2.95 &$HANN$ \\
Si\,{\sc I} &  6125.03 &  5.61 & $-$1.540 &   42.1  &  8.62 &$REDH$  &Y\,{\sc II} &  6613.73 &  1.75 & $-$1.110 &72.3	&  2.92 &$HANN$ \\
S\,{\sc I}  &  4694.12 &  6.52 & $-$1.713 &   65.2  &  7.98 &$PODK$  &Zr\,{\sc II}&  4208.98 &  0.71 & $-$0.510 &syn	&  3.75 &$LJUN$ \\
S\,{\sc I}  &  4695.45 &  6.52 & $-$1.871 &   42.8  &  7.92 &$PODK$  &Zr\,{\sc II}&  4211.87 &  0.53 & $-$1.040 &syn	&  3.55 &$LJUN$ \\
S\,{\sc I}  &  6743.57 &  7.86 & $-$1.065 &   63.5  &  8.32 &$PODK$  &Zr\,{\sc II}&  4370.95 &  1.21 & $-$0.770 &syn	&  3.40 &$LJUN$ \\
S\,{\sc I}  &  6748.78 &  7.87 & $-$0.638 &  102.6  &  8.20 &$PODK$  &Zr\,{\sc II}&  4379.78 &  1.53 & $-$0.356 &syn	&  3.40 &$KRCZ$ \\
S\,{\sc I}  &  6757.19 &  7.87 & $-$0.351 &  124.1  &  8.04 &$PODK$  &Zr\,{\sc II}&  4440.45 &  1.21 & $-$1.040 &164.8  &  3.45 &$LJUN$ \\
S\,{\sc I}  &10\,459.41&  6.86 &    0.030 &  syn    &  8.32 &$SCOT$  &Zr\,{\sc II}&  4496.97 &  0.71 & $-$1.540 &syn	&  3.55 &$THEV$ \\
Ca\,{\sc I} &  4425.44 &  1.88 & $-$0.358 &  syn    &  6.86 &$NIST$  &Zr\,{\sc II}&  4661.79 &  2.66 & $-$0.620 &69.1	&  3.58 &$THEV$ \\
Ca\,{\sc I} &  4585.87 &  2.52 & $-$0.187 &   57.7  &  6.58 &$NIST$  &Ba\,{\sc II}&  5853.69 &  0.60 & $-$1.010 &syn	&  2.65 &$MCMW$ \\
Ca\,{\sc I} &  5581.98 &  2.52 & $-$0.710 & 59.5    &  7.06 &$NIST$  &La\,{\sc II}&  4333.76 &  0.17 & $-$0.060 &syn	&  1.57 &$L001$ \\
Ca\,{\sc I} &  5601.29 &  2.53 & $-$0.690 & 51.7    &  6.97 &$NIST$  &La\,{\sc II}&  4322.50 &  0.17 & $-$0.930 &syn	&\le1.72&$L001$ \\
Ca\,{\sc I} &  5857.46 &  2.93 &    0.230 &127.3    &  6.90 &$NIST$  &Nd\,{\sc II}&  4061.08 &  0.47 &    0.550 &syn	&  2.10 &$DEHG$ \\
Ca\,{\sc I} &  6122.23 &  1.89 & $-$0.315 &180.1    &  6.89 &$NIST$  &Nd\,{\sc II}&  5319.81 &  0.55 & $-$0.140 &syn	&  2.05 &$DEHG$ \\
Ca\,{\sc I} &  6162.18 &  1.90 & $-$0.089 &234.3    &  6.98 &$NIST$  &Eu\,{\sc II}&  4129.70 &  0.00 &    0.220 &syn	&  1.65 &$L001$ \\
Ca\,{\sc I} &  6449.82 &  2.52 & $-$0.552 &58.7     &  6.88 &$NIST$  &Eu\,{\sc II}&  6645.06 &  1.38 &    1.709 &syn	&  1.53 &$L001$ \\
Sc\,{\sc II}&  4420.66 &  0.62 & $-$2.270 &197.6    &  4.05 &$NIST$  &Eu\,{\sc II}&  7426.57 &  1.28 &    0.211 &syn	&  1.76 &$L001$  \\
\hline 
\end{array}
$$
\begin{list}{}{}
\item
{\bf (a)} References for the adopted $gf$ values: WFD-Wiese, Fuhr, Deters (1996); NIST-Atomic
Spectra Database {(http://physics.nist.gov/PhysRefData/ASD)}; KRCZ-KURUCZ Atomic Spectra Database
{(http://www.pmp.uni-hannover.de/projekte/kurucz/)}; THEV-Th{\'e}venin (1989); KELP-Kelleher \&
Podobedova (2008a); REDH-Reddy \& Hrivnak (1999); PODK-Podobedova,Kelleher, Wiese (2009);
SOLS-Sobeck, Lawler, Sneden(2007); SCOT-Scott et al. (2015); BIEG-Bi\'{e}mont \& Godefroid (1980);
HANN-Hannoford et al. (1982); LJUN-Ljung et al. (2006); MCMW-Mcwilliam et al. (1998); L001-Lawler,
Bonvallet, Sneden (2001a); DEHG-Den Hartog et al. (2003); L013-Wood et al. (2013).
\end{list}
\end{table*}

\begin{table*}
\caption{Fe\,{\sc i} and Fe\,{\sc ii} lines used in the analysis of McDonald and SAO spectra. Abundances for individual lines are
those obtained for the 2009 May 9 spectrum and a model of $T_{\rm eff}=7350$ K, $\log$ g = 0.64
and $\xi$=6.6 km s$^{\rm -1}$.}
\label{}
$$
        \begin{array}{@{}l|crrrc||l|crrrc}
\hline           \hline
 $Species$&\lambda &  $LEP$&\log(gf)& $EW$&\log\epsilon(Fe)&$Species$&\lambda &$LEP$&\log(gf)& $EW$&\log\epsilon(Fe)\\
\cline{2-3}
\cline{5-6}
\cline{8-9}
\cline{11-12}
          &($\AA$) & ($eV$) &        & (m$\AA$)&(dex)     &	  & ($\AA$)&($eV$)& 	&(m$\AA$)&($dex$)       \\
\hline           \hline
Fe\,{\sc I}  & 4009.71 &  2.22 &  -1.25  &   192.1  &  7.88 & Fe\,{\sc I}  & 6024.07 & 4.55 &  -0.06 &    96.8  & 7.91 \\ 
Fe\,{\sc I}  & 4107.49 &  2.83 &  -0.88  &   170.2  &  7.87 & Fe\,{\sc I}  & 6027.06 & 4.08 &  -1.09 &    43.5  & 8.16 \\ 
Fe\,{\sc I}  & 4213.65 &  2.84 &  -1.25  &    87.3  &  7.81 & Fe\,{\sc I}  & 6065.49 & 2.61 &  -1.53 &    70.0  & 7.68 \\ 
Fe\,{\sc I}  & 4484.23 &  3.60 &  -0.86  &   102.8  &  8.07 & Fe\,{\sc I}  & 6393.61 & 2.43 &  -1.58 &    87.3  & 7.70 \\ 
Fe\,{\sc I}  & 4602.95 &  1.49 &  -2.22  &   151.2  &  8.01 & Fe\,{\sc I}  & 6400.01 & 3.60 &  -0.29 &   122.9  & 7.54 \\ 
Fe\,{\sc I}  & 4643.47 &  3.65 &  -1.15  &    47.8  &  7.98 & Fe\,{\sc I}  & 6411.66 & 3.65 &  -0.72 &    98.6  & 7.88 \\ 
Fe\,{\sc I}  & 5090.77 &  4.26 &  -0.44  &   101.1  &  8.11 & Fe\,{\sc I}  & 6592.91 & 2.73 &  -1.47 &    71.7  & 7.72 \\
Fe\,{\sc I}  & 5202.34 &  2.18 &  -1.84  &   144.2  &  8.10 & Fe\,{\sc I}  & 6841.34 & 4.61 &  -0.60 &    50.3  & 8.15 \\
Fe\,{\sc I}  & 5353.37 &  4.10 &  -0.68  &    77.2  &  8.07 & Fe\,{\sc II} & 4893.82 & 2.83 &  -4.27 &   238.5  & 8.06 \\ 
Fe\,{\sc I}  & 5364.88 &  4.45 &   0.23  &   157.0  &  7.86 & Fe\,{\sc II} & 5427.80 & 6.72 &  -1.58 &    91.1  & 7.58 \\ 
Fe\,{\sc I}  & 5367.48 &  4.42 &   0.44  &   190.7  &  7.78 & Fe\,{\sc II} & 5813.67 & 5.57 &  -2.75 &    86.4  & 7.89 \\ 
Fe\,{\sc I}  & 5373.71 &  4.47 &  -0.84  &    10.1  &  7.54 & Fe\,{\sc II} & 5823.18 & 5.57 &  -2.99 &    67.9  & 7.99 \\ 
Fe\,{\sc I}  & 5383.38 &  4.31 &   0.65  &   259.7  &  7.80 & Fe\,{\sc II} & 5991.38 & 3.15 &  -3.65 &   223.0  & 7.59 \\ 
Fe\,{\sc I}  & 5393.18 &  3.24 &  -0.72  &   133.9  &  7.76 & Fe\,{\sc II} & 6113.33 & 3.22 &  -4.23 &   137.7  & 7.84 \\ 
Fe\,{\sc I}  & 5434.53 &  1.01 &  -2.12  &   234.4  &  7.82 & Fe\,{\sc II} & 6116.06 & 3.23 &  -4.47 &    87.3  & 7.82 \\
Fe\,{\sc I}  & 5569.63 &  3.42 &  -0.49  &   149.4  &  7.74 & Fe\,{\sc II} & 6129.73 & 3.20 &  -4.74 &   111.0  & 8.20 \\ 
Fe\,{\sc I}  & 5572.85 &  3.40 &  -0.28  &   209.4  &  7.78 & Fe\,{\sc II} & 6179.40 & 5.57 &  -2.80 &    69.4  & 7.83 \\ 
Fe\,{\sc I}  & 5615.66 &  3.33 &   0.05  &   270.5  &  7.67 & Fe\,{\sc II} & 6331.95 & 6.22 &  -2.07 &   124.7  & 7.94 \\
Fe\,{\sc I}  & 5816.38 &  4.55 &  -0.60  &    35.0  &  7.93 & Fe\,{\sc II} & 6446.40 & 6.22 &  -2.08 &    89.4  & 7.75 \\ 
Fe\,{\sc I}  & 6020.17 &  4.61 &  -0.21  &    56.9  &  7.82 &		   &	     &      &	     &  	&      \\
\hline 
\hline
\end{array}
$$
\end{table*}

\subsection{Radial Velocities}

Radial velocities for the McDonald spectra were measured
using cross-correlation against a spectrum of Arcturus (Hinkle et al.
2000). The spectral range used for the cross-correlation was 5300-5500\AA. 
As a check, we also derived radial velocities from the central wavelengths
of several unblended Fe\,{\sc i} lines and the laboratory wavelengths
(Nave et al. 1994). For a given spectrum, the two measurement techniques
agreed to within about 0.4 km s$^{-1}$. These velocities are based on the
wavelength scale derived from exposures of a ThAr hollow cathode lamp
observed during the same night but rarely either immediately before or
after the HD 179821 exposure. To correct for a possible offset between
stellar and ThAr lamp exposures, velocities were derived from telluric
(H$_2$O and O$_2$) lines providing a correction of 0.1-0.4 km s$^{-1}$.

Procedures for obtaining radial velocities from the SAO spectra are
described by Klochkova et al. (2008). The list of selected lines is
taken from the line list for the post-AGB F supergiant HD 56126 (Klochkova
et al. 2007) and again the stability of the spectrograph was
checked using telluric ([O\,{\sc i}], H$_2$O and O$_2$) lines. An
accuracy of 0.1-0.2 km s$^{-1}$ is achieved. 

The heliocentric velocities are listed in Table 2. A variable
radial velocity is evident with a peak-to-peak amplitude of about
15 km s$^{-1}$. The central velocity and amplitude are consistent
with measurements in the literature from about 10 years earlier. The mean velocity of
85.8$\pm$0.8 km s$^{\rm -1}$ (V$_{\rm LSR}=$102.3$\pm$0.8 km s$^{\rm -1}$)from
Table 2 is in good agreement with the systemic LSR velocity (100 km s$^{\rm -1}$) obtained from CO
observations of the expanding circumstellar shell (Zuckerman \& Dyck 1986; Likkel et al. 1987;
van der Veen et al. 1993; Fong et al. 2006; Castro-Carrizo et al. 2007).

Photometrically, HD 179821 is a semi-regular variable. Arkhipova et al. (2001; see also Le
Coroller et al. 2003) from photometric observations of the star between 1994-1999, found low amplitude variations
($\Delta V \simeq 0.05 - 0.20$ magnitudes) and reported the periods for
fundamental pulsation (P$_{\rm 0}$) and for the first overtone (P$_{\rm 1}$) as 205$^{\rm
d}$ and 142$^{\rm d}$, respectively. Revised periods from UBV photometry from 2000 to 2008 by Arkhipova et al (2009) were P$_{\rm 0}$ = 203$\pm5$$^{\rm
d}$ and P$_{\rm 1}$ = 141$\pm5$$^{\rm d}$. The ratio P$_{\rm 1}$/P$_{\rm 0}$ is
$\approx$0.7, a value of P$_{\rm
1}$/P$_{\rm 0}$ is close to 0.705 in classical cepheids (Stobie 1977). Our radial velocity
data are too sparse to confirm these periods but the pulsations are surely responsible for
radial velocity variations. 

A characteristic of all luminous supergiants is the large
width of photospheric lines. HD\,179821 is no exception; the
line profiles provide a macroturbulent velocity of 22$\pm2$ km s$^{-1}$
(see below). Although computation of the synthetic spectrum
assumes that the macroturbulence is uniform across the atmosphere, 
the atmosphere may be more realistically spotted with a few large
convection cells, as suggested by Schwarzschild (1975) for red supergiants. The appearance and
disappearance of a cell among the few populating the visible surface
can lead to a profile variation and a radial
velocity change.   

\subsection{Photospheric line profile variations}

Asymmetric line profiles are almost a common property of the spectra over all epochs and also
reported by Kipper (2008). In the 2009 McDonald spectrum, the lines have almost symmetric
profiles and it was selected
for the abundance analysis.

The shapes of the line profiles vary with their strengths. For several relatively high
excitation lines, the broadening of the line is seen to occur mainly at the long
wavelength side of the line profiles and the short wavelength side of the line profiles
remains almost static. On the other hand, for the low excitation lines, the broadening is
observed to occur mainly at the short wavelength side of the line profiles.

Quantitative representation of the observed line profile variations are presented in Tables A1
and A2.
In these tables, the measured EWs of neutral and ionized iron lines from the spectra over all epochs 
are listed and indicate notable changes in the spectra for 2008,
2010, and 2013 epochs. These changes seem to correlate with the episodic variation detected in the
H$\alpha$ profile, for instance, for the 2010 May and July spectra.  
\begin{table}
\caption{Solar abundances obtained by employing the solar model atmosphere from
Castelli \& Kurucz (2004) compared to the photospheric abundances from Asplund et al.
(2009). The abundances presented in bold typeface are measured by synthesis while
remaining elemental abundances were calculated using the line EWs.}
\label{}
$$
        \begin{array}{@{}l@{}|c|@{}c|c|@{}c}
           \hline
           \hline

          &This\,work       & & Asplund &        \\
\cline{2-2}
\cline{4-5}
 $Species$&\log\epsilon{_\odot}(X)& \,No. &\log\epsilon{_\odot}(X) & \,\Delta\log\epsilon{_\odot}(X)^{*}\\
\cline{2-2}
 \cline{4-5}

          &(dex)       & & (dex) & (dex)        \\
           \hline
\hline
C\,{\sc I}    &8.45\pm0.15       &7  & 8.43\pm0.05 & 0.02     \\
N\,{\sc I}    &8.08\pm0.08 	 &2  & 7.83\pm0.05 & 0.25     \\
O\,{\sc I}    &8.76\pm0.21 	 &2  & 8.69\pm0.05 & 0.07     \\
Na\,{\sc I}   &6.16\pm0.01 	 &2  & 6.24\pm0.04 & $-$0.08  \\
Mg\,{\sc I}   &7.60\pm0.00 	 &1  & 7.60\pm0.04 & 0.00     \\
Mg\,{\sc II}  &{\bf 7.54\pm0.01} &2  & 7.60\pm0.04 & $-$0.06  \\
A\,{\sc I}    &{\bf 6.42\pm0.00} &2  & 6.45\pm0.03 & $-$0.03  \\
Si\,{\sc I}   &7.55\pm0.07	 &6  & 7.51\pm0.03 & 0.04     \\
S\,{\sc I}    &7.12\pm0.04	 &4  & 7.12\pm0.03 & 0.00     \\
Ca\,{\sc I}   &6.20\pm0.23 	 &8  & 6.34\pm0.04 & $-$0.14  \\
Sc\,{\sc II}  &3.11\pm0.09 	 &4  & 3.15\pm0.04 & $-0.04$  \\
Ti\,{\sc II}  &{\bf 5.01\pm0.15} &5  & 4.95\pm0.05 & 0.06     \\
Cr\,{\sc I}   &5.54\pm0.10 	 &3  & 5.64\pm0.04 & $-$0.10  \\
Cr\,{\sc II}  &5.68\pm0.06 	 &4  & 5.64\pm0.04 & 0.04     \\
Mn\,{\sc I}   &5.67\pm0.05 	 &3  & 5.43\pm0.04 & 0.24     \\
Fe\,{\sc I}   &7.40\pm0.11	 &24 & 7.50\pm0.04 & $-$0.10  \\
Fe\,{\sc II}  &7.50\pm0.08 	 &9  & 7.50\pm0.04 & 0.00     \\
Ni\,{\sc I}   &6.22\pm0.19 	 &7  & 6.22\pm0.04 & 0.00     \\
Zn\,{\sc I}   &4.66\pm0.26       &2  & 4.56\pm0.05 & 0.10     \\
Y\,{\sc II}   &2.35\pm0.19       &3  & 2.21\pm0.05 & 0.14     \\ 
Zr\,{\sc II}  &2.71\pm0.10       &2  & 2.58\pm0.04 & 0.13     \\
Ba\,{\sc II}  &{\bf 2.13\pm0.00} &1  & 2.18\pm0.09 & $-$0.05  \\
La\,{\sc II}  &{\bf 1.03\pm0.00} &1  & 1.10\pm0.04 & $-$0.07  \\
Nd\,{\sc II}  &{\bf 1.37\pm0.00} &1  & 1.42\pm0.04 & $-$0.05  \\
Eu\,{\sc II}  &{\bf 0.43\pm0.02} &3  & 0.52\pm0.04 & $-$0.09  \\
\hline 
\hline
\end{array}
$$
\begin{list}{}{}
\item \hskip 0.5 cm {\bf (*):}  $\Delta\log\epsilon{_\odot}(X)= \log\epsilon{_\odot}(X)_{\rm This\,work} -
\log\epsilon{_\odot}(X)_{\rm Asplund}$
\end{list}
\end{table}

\subsection{Interstellar lines}

Interstellar lines may prove relevant to the interpretation of HD 179821
in two ways: their radial velocities in conjunction with a model of
Galactic kinematics place constraints on the the star's distance and the
equivalent widths of lines with an appropriate calibration may provide an
estimate of the interstellar reddening, an essential ingredient in
assessing the star's luminosity. For HD 179821 with its circumstellar
gas and dust, there is the possibility of contamination of interstellar
lines by circumstellar components, especially in the case of the Na D lines with their
complex profile.

Diffuse interstellar bands (DIBs) cross the optical spectrum, as noted previously, with Kipper (2008)
providing the most complete discussion. The DIBs are assumed to be of interstellar origin. Mean
radial velocities from 4 to 17 DIBs are given in Table 2. The rest wavelengths for DIBs are provided
by Hobbs et al. (2008). As expected, the mean velocity does not vary from spectrum to spectrum. The
mean is $-8.5\pm0.5$ km s$^{-1}$ (over all epochs), a value consistent with that obtained by Kipper
(2008). The corresponding LSR velocity is $8.1\pm0.5$ km s$^{-1}$. An LSR velocity of 8--10 km s$^{\rm -1}$ is
observed in the interstellar medium (ISM) in the direction of the star in the Galaxy (Brand \&
Blitz 1993). This permits us to conclude that the DIBs detected in spectra of HD\,179821
are, indeed, formed in the ISM, however, the negative velocity measured for this principal ISM cloud
can not be used to estimate the distance of the star. 

Kipper (2008) discusses two calibrations of the reddening
$E(B-V)$ versus equivalent widths of the measured DIBs. For the
$-8.5$ km s$^{-1}$ leading component, we obtain $E(B-V) \simeq
1.0$, a value in fair agreement with the estimate of 0.7 from the
broad-band colors (Arkhipova et al. 2009).

 \begin{table*}
\begin{center}
    \caption[]{Abundances of the observed species for HD\,179821 are presented
    for the 9 May 2009 McDonald spectrum and the model atmospheres of $T_{\rm eff} = 7350$ K, $\log$ g = 0.64, $\xi$ = 6.6 km s$^{\rm -1}$.}
       \label{}
   $$
       \begin{array}{l@{}||cccc|cc|cc||c}
          \hline
 $Element$       &\log\epsilon(X)&\sigma_{\rm line}&\sigma_{\rm abs} &$N$ &$[X/H]$ &\sigma_{\rm [X/H]}&$[X/Fe]$&\sigma_{\rm [X/Fe]}& \log\epsilon{_\odot}(X)\\     
\cline{2-4}
\cline{6-7}
\cline{8-9}
                 &$(dex)$&$(dex)$&$(dex)$& &$(dex)$& $(dex)$ & $(dex)$ &$(dex)$& $(dex)$	  \\ 
\hline
 C$\,{\sc i}$    & 8.79&0.19 &0.18 &10   & 0.34&0.24 &$-$0.19&0.30 & 8.45\pm0.15  \\
 N$\,{\sc i}$    & 9.48&0.14 &0.24 & 3   & 1.40&0.16 &   0.87&0.24 & 8.08\pm0.08  \\ 
 O$\,{\sc i}$    & 9.04&0.06 &0.06 & 2   & 0.28&0.22 &$-$0.25&0.28 & 8.76\pm0.21  \\
 Na$\,{\sc i}$   & 7.67&0.01 &0.31 & 2   & 1.51&0.01 &   0.98&0.18 & 6.16\pm0.01  \\
 Mg$\,{\sc i}$   & 8.32&0.08 &0.32 & 4   & 0.72&0.08 &   0.19&0.20 & 7.60\pm0.00  \\
 Mg$\,{\sc ii}$  & 8.40&0.10 &0.23 & 3   & 0.86&0.10 &   0.33&0.21 & 7.54\pm0.01  \\ 
 Al$\,{\sc i}$   & 6.78&0.01 &0.20 & 2   & 0.36&0.01 &$-$0.17&0.18 & 6.42\pm0.00  \\
 Si$\,{\sc i}$   & 8.58&0.15 &0.23 & 6   & 1.03&0.17 &   0.50&0.25 & 7.55\pm0.07  \\
 S$\,{\sc i}$    & 8.13&0.17 &0.21 & 6   & 1.01&0.17 &   0.48&0.25 & 7.12\pm0.04  \\
 Ca$\,{\sc i}$   & 6.89&0.14 &0.33 & 8   & 0.69&0.27 &   0.16&0.32 & 6.20\pm0.23  \\
 Sc$\,{\sc ii}$  & 3.90&0.18 &0.17 & 4   & 0.79&0.20 &   0.26&0.27 & 3.11\pm0.09  \\
 Ti$\,{\sc ii}$  & 5.80&0.13 &0.21 & 8   & 0.79&0.20 &   0.26&0.27 & 5.01\pm0.15  \\
 Cr$\,{\sc i}$   & 6.04&0.04 &0.29 & 3   & 0.50&0.11 &$-$0.03&0.21 & 5.54\pm0.10  \\
 Cr$\,{\sc ii}$  & 6.16&0.04 &0.15 & 4   & 0.48&0.07 &$-$0.05&0.19 & 5.68\pm0.06  \\
 Mn$\,{\sc i}$   & 5.86&0.04 &0.27 & 3   & 0.19&0.06 &$-$0.34&0.19 & 5.67\pm0.05  \\
 Fe$\,{\sc i}$   & 7.93&0.14 &0.26 &28   & 0.53&0.18 &   0.00&0.25 & 7.40\pm0.11  \\
 Fe$\,{\sc ii}$  & 7.93&0.19 &0.09 &11   & 0.43&0.21 &$-$0.10&0.28 & 7.50\pm0.08  \\
 Ni$\,{\sc i} $  & 7.03&0.14 &0.25 & 7   & 0.81&0.24 &   0.28&0.30 & 6.22\pm0.19  \\
 Ni$\,{\sc ii}$  & 7.15&0.00 &0.19 & 1   & 0.93&0.00 &   0.40&0.18 & ...	  \\
 Zn$\,{\sc i}$   & 5.08&0.15 &0.26 & 2   & 0.42&0.30 &$-$0.11&0.35 & 4.66\pm0.26  \\
 Y $\,{\sc ii}$  & 2.84&0.11 &0.21 & 4   & 0.49&0.22 &$-$0.04&0.28 & 2.35\pm0.19  \\
 Zr$\,{\sc ii}$  & 3.53&0.12 &0.15 & 7   & 0.82&0.16 &   0.29&0.24 & 2.71\pm0.10  \\
 Ba$\,{\sc ii}$  & 2.65&0.00 &0.29 & 1   & 0.52&0.00 &$-$0.01&0.18 & 2.13\pm0.00  \\
 La$\,{\sc ii}$  & 1.65&0.11 &0.36 & 2   & 0.62&0.11 &   0.09&0.21 & 1.03\pm0.00  \\
 Nd$\,{\sc ii}$  & 2.08&0.04 &0.27 & 2   & 0.71&0.04 &   0.18&0.18 & 1.37\pm0.00  \\
 Eu$\,{\sc ii}$  & 1.65&0.12 &0.20 & 3   & 1.22&0.12 &   0.69&0.22 & 0.43\pm0.02  \\
\hline
       \end{array}
   $$
\end{center}
 \end{table*}

\begin{table}
\begin{center}
\caption{Model atmosphere parameters from McDonald and SAO spectra.}
$$
\begin{array}{l|l|c|c|c}
\hline
               & \multicolumn{4}{c}{This\,Work}\\
\cline{2-5}
$Obs. Period$  &\hskip 0.5cm\Teff &\logg  & $[Fe/H]$  &\xi \\
\cline{2-5}
              & \hskip 0.5cm (\kelvin)  & ($c.g.s$)  & ($dex$)    & (km s^{-1})  \\
\hline
 21 Apr 2008 & 7300\pm100 &  0.50\pm0.25  &  0.25\pm0.15  &  9.1\pm1.5 \\
 13 Jun 2008 & 7300\pm100 &  0.50\pm0.25  &  0.38\pm0.15  &  7.2\pm1.4 \\
 11 Jul 2008 & 7300\pm120 &  0.55\pm0.30  &  0.30\pm0.17  &  6.8\pm1.3 \\
 10 Aug 2008 & 7250\pm130 &  0.50\pm0.27  &  0.20\pm0.19  &  7.7\pm1.7 \\
 17 Aug 2008 & 7350\pm80  &  0.53\pm0.18  &  0.50\pm0.12  &  5.4\pm1.3 \\
 17 Sep 2008 & 7400\pm100 &  0.50\pm0.13  &  0.20\pm0.14  &  7.6\pm0.9 \\
 14 Apr 2009 & 7400\pm150 &  0.60\pm0.23  &  0.30\pm0.17  &  7.8\pm1.2 \\
 09 May 2009 & 7350\pm200 &  0.64\pm0.30  &  0.43\pm0.19  &  6.6\pm1.6 \\
 07 Nov 2009 & 7220\pm80  &  0.50\pm0.14  &  0.00\pm0.12  &  6.7\pm1.2 \\
 21 Nov 2009 & 7400\pm120 &  0.60\pm0.26  &  0.35\pm0.17  &  8.7\pm1.4 \\
 22 May 2010 & 7400\pm110 &  0.85\pm0.41  &  0.45\pm0.18  &  8.9\pm1.8 \\
 03 Jun 2010 & 7200\pm100 &  0.55\pm0.30  &  0.25\pm0.15  &  7.4\pm1.1 \\
 30 Jul 2010 & 7400\pm80  &  0.50\pm0.65  &  0.47\pm0.45  &  6.0\pm1.3 \\
 24 Sep 2010 & 7300\pm100 &  0.50\pm0.20  &  0.40\pm0.14  &  6.4\pm1.4 \\
 17 Nov 2010 & 7150\pm120 &  0.50\pm0.27  &  0.27\pm0.14  &  7.7\pm1.2 \\
 16 May 2011 & 7300\pm130 &  0.50\pm0.35  &  0.33\pm0.19  &  7.9\pm1.5 \\
 27 Aug 2013 & 7360\pm130 &  0.72\pm0.40  &  0.50\pm0.18  &  8.1\pm1.6 \\
\cline{1-5}
 {\bf Mean}        &7316\pm77   &  0.56\pm0.10  & 0.33\pm0.13   &  7.4\pm1.0 \\
\hline \hline
\end{array}
$$
\end{center} 
\end{table}

Profiles of the two  NaD lines were illustrated by Za\v{c}s et al. (1996),
Reddy \& Hrivnak (1999), and Kipper (2008) -- see Fig. 3. The predicted photospheric Na D1  profile for 2009 May 9 
is located between absorption components 4 and 5. Components 4 and 5 are seen to vary in intensity
over all epochs. Varying intensity of these components may be an indication for a circumstellar
component.

Velocities of the Na D components are constant to within the measurement errors and
certainly independent of the varying photospheric velocity. Component 1 has a
mean velocity of $-10.5\pm0.8$ km s$^{-1}$. Given the
uncertainty about the rest wavelengths of the DIBs, this represents satisfactory agreement
between the NaD component and the principal DIB component. Components 2 to 5 may also be of interstellar origin.
The Galactic rotation model of Brand \& Blitz (1993) provides a crude estimate of distance to the component 2 as
$\approx$ 1.5 kpc and $\ge$6.9 kpc for component 5.

Kipper (2008) using a calibration by Munari \& Zwitter (1997) estimated the
reddening provided by each Na D component. If contributions from all
components are summed, the total is $E(B-V) \simeq 0.8$, a value in agreement
with estimates from the broad band colors and the DIBs. An identification of
one or more of the Na D components with the circumstellar material could change
this estimate. 

 \section{The abundance analysis}
All abundance analyses were performed using standard {\sc ATLAS9} LTE model atmospheres (Castelli \&
Kurucz 2004)\footnote{http://kurucz.harvard.edu/grids.html} and a recent version of the LTE line
analysis {\sc MOOG} code (Sneden 1973)\footnote{http://www.as.utexas.edu/chris/moog.html}. All calculations
assume a normal He abundance (He/H=0.085). In the following subsections, we discuss the adopted line
list, the derivation of the model atmosphere parameters, and comment briefly on the obtained
abundances. Then, we compare our results with previously published abundances before presenting our
conclusions in Section 5 about HD 179821's composition and its evolutionary status.

 \begin{table}
\caption[]{Sensitivity of the derived abundances to the uncertainties of $\Delta \Teff = +200$ \kelvin,
$\Delta\log$ g = +0.3, and $\Delta\xi = \pm$ 1.6 in the model atmosphere parameters for $\Teff = 7350$ \kelvin, $\log$ g = 0.64, and $\xi = \pm$ 6.6.}
       \label{}
   $$
       \begin{array}{@{}l||c|@{}c|@{}c|@{}c}
          \hline
          \hline
         &                 \multicolumn{4}{c}{\Delta log\,\epsilon}             \\
\cline{2-5}
$Species$         &  \Delta T_{\rm eff} & \,\Delta log\,g & \Delta\,\xi  &\Delta\,\xi   \\
\cline{2-5}

  &    $+200$       &  $+0.3$      &  $+1.6$     & $-1.6$ \\
\cline{2-5}
  &       ($K$)     &   ($cgs$)    & ($km$\,s^{-1})&($km$\,s^{-1}) \\
          \hline

 C$\,{\sc i}$    &+0.15  &$-$0.09 &$-$0.03&+0.05  \\
 N$\,{\sc i}$    &$-$0.01&$-$0.03 &$-$0.24&+0.37  \\ 
 O$\,{\sc i}$    &$-$0.04&0.00    &$-$0.05&+0.04  \\
 Na$\,{\sc i}$   &+0.31 &+0.01   &$-$0.04 &+0.45  \\
 Mg$\,{\sc i}$   &+0.20 &$-$0.13 &$-$0.21 &+0.42  \\
 Mg$\,{\sc ii}$  &+0.02 &+0.05   &$-$0.22 &+0.41  \\ 
 Al$\,{\sc i}$   &+0.17 &$-$0.10 &0.00    &+0.02  \\ 
 Si$\,{\sc i}$   &+0.20 &$-$0.12 &$-$0.01 &+0.03  \\
 S$\,{\sc i}$    &+0.15 &$-$0.14 &$-$0.06 &+0.02  \\
 Ca$\,{\sc i}$   &+0.29 &$-$0.16 &$-$0.04 &+0.11  \\
 Sc$\,{\sc ii}$  &+0.12 &$-$0.04 &$-$0.12 &+0.13  \\
 Ti$\,{\sc ii}$  &+0.15 &+0.03   &$-$0.15 &+0.29  \\
 Cr$\,{\sc i}$   &+0.27 &$-$0.10 &0.00    &+0.07  \\
 Cr$\,{\sc ii}$  &+0.11 &+0.03   &$-$0.10 &+0.26  \\
 Mn$\,{\sc i}$   &+0.24 &$-$0.12 &$-$0.02 &+0.07  \\
 Fe$\,{\sc i}$   &+0.22 &$-$0.13 &$-$0.07 &+0.12  \\
 Fe$\,{\sc ii}$  &+0.07 &+0.02   &$-$0.06 &+0.11  \\
 Ni$\,{\sc i} $  &+0.20 &$-$0.14 &$-$0.04 &+0.02  \\
 Ni$\,{\sc ii}$  &+0.10 &+0.05   &$-$0.15 &+0.40  \\
 Zn$\,{\sc i} $  &+0.23 &$-$0.12 &$-$0.02 &+0.05  \\
 Y $\,{\sc ii}$  &+0.21 &$-$0.01 &$-$0.04 &+0.12  \\
 Zr$\,{\sc ii} $ &+0.14 &$-$0.01 &$-$0.06 &+0.09  \\
 Ba$\,{\sc ii}$  &+0.26 &$-$0.12 &$-$0.04 &+0.06  \\
 La$\,{\sc ii}$  &+0.35 &$-$0.02 &$-$0.07 &+0.05  \\
 Nd$\,{\sc ii}$  &+0.25 &$-$0.10 &$-$0.05 &0.00   \\
 Eu$\,{\sc ii}$  &+0.17 &$-$0.09 &$-$0.05 &+0.03  \\
\hline
\hline
       \end{array}
   $$
 \end{table}

\begin{figure}
\begin{center}
\includegraphics[width=85mm,height=70mm,angle=0]{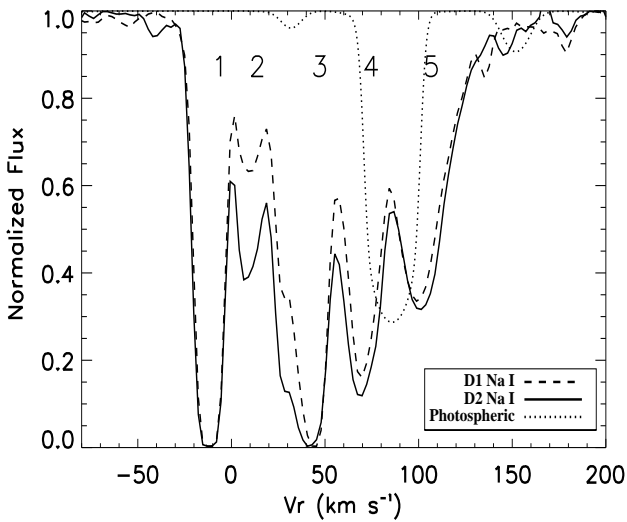}
\caption{The profile of the D$_{2}$ (solid) and D$_{1}$ (dotted) lines of Na I in the McDonald 2009
May 9 spectrum of HD\,179821. Telluric H$_{2}$O lines
have not been removed. Photospheric profile indicated with dotted line is computed for \Teff = 7350 K and log g = 0.64.}
\end{center}
\end{figure}

\subsection{The line list}
An essential prerequisite for the abundance analysis of HD\,179821 is a set of securely identified
lines with reliable atomic data. Our line lists were generated from a systematic search for unblended 
lines. In about 30 per cent of the accepted lines, spectrum synthesis was preferred to a direct
estimate of equivalent width; known contaminating transitions in one or both wings were thus taken
into account. Our final list covers 21 elements and 141 lines over the spectra range from about 4000
\AA\ to 10\,460 \AA. The list of lines for elements other than Fe is provided in Table 3
with $gf$-values taken from the literature with references supplied in 
the notes to Table 3.
Chosen lines of Fe\,{\sc i} and Fe\,{\sc ii} are listed in Table 4 with
$gf$-values taken from the critical compilationby Fuhr \& Wiese (2006). 

\noindent As a check on the line list and especially on the selection of $gf$-values, we have derived solar
abundances using our list. Our lines were measured off the solar flux atlas of Kurucz et al.'s (1984)
and analyzed with the solar model atmosphere from the Castelli \& Kurucz (2004) grid for $T_{\rm eff}
=$ 5777 K and $\logg  =$ 4.44. Our analysis gave a microturbulence of 0.87 km s$^{-1}$ and the
abundances in Table 5. The estimated solar abundances are compared with  those from Asplund et al.
(2009) in their critical review. The agreement is good but for two elements, e.g., N and Mn where our
abundances are 0.25 dex greater. Perfect agreement is not expected because line lists, selection of
$gf$-values and model atmospheres differ. In referencing stellar abundances to solar values, we use
our solar abundances. 

\begin{figure*} \begin{center}
\includegraphics[width=180mm,height=125mm,angle=0]{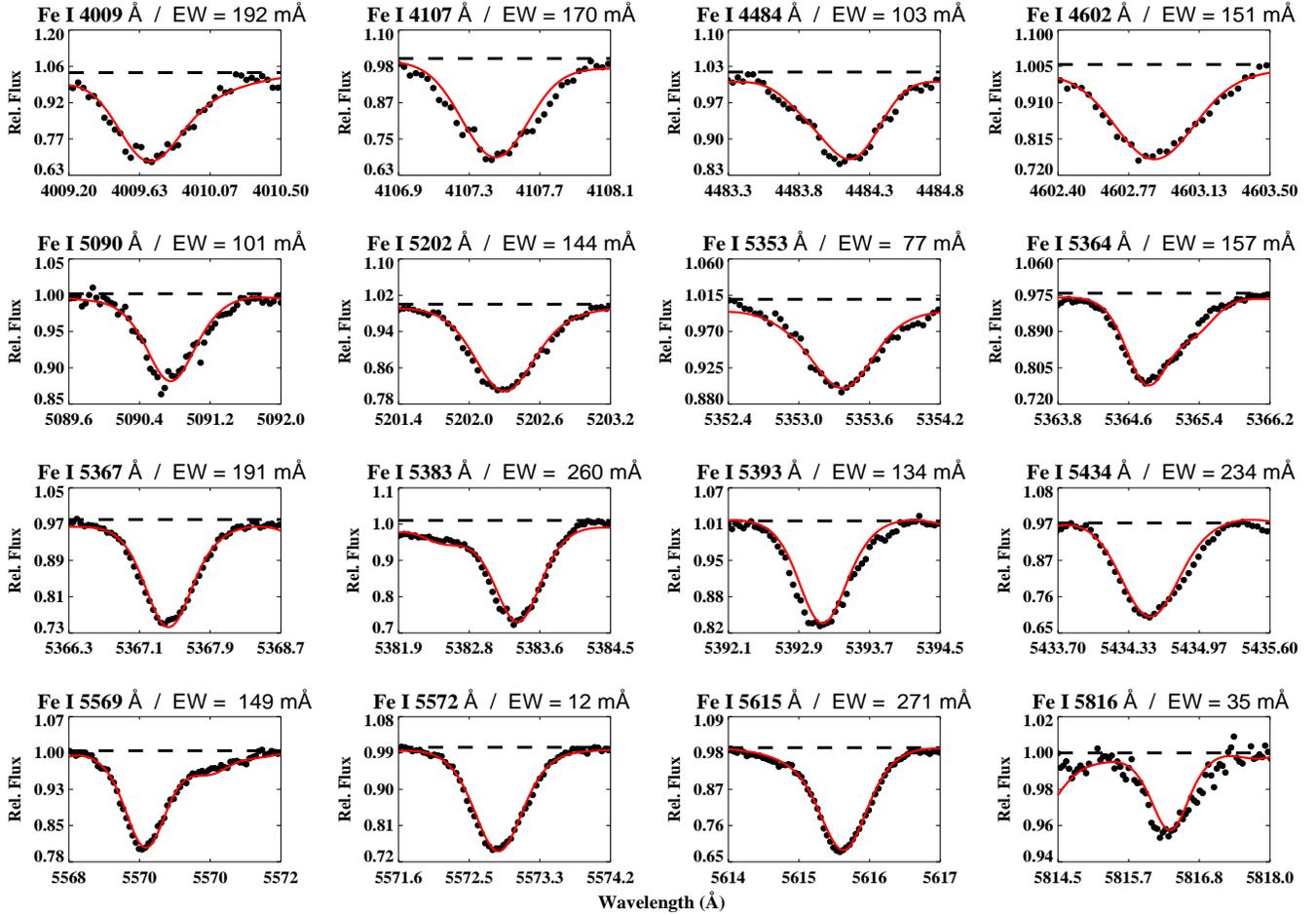} 
\caption{The observed (filled circles) and computed (full red line) line profiles for neutral Fe lines used for model parameter determination from the
McDonald spectrum for 2009 May 9. Their wavelengths and measured equivalent widths (EW) are
indicated at the top of each panel. The computed profiles show synthetic spectra for the abundances listed in
Table 4 and blending lines included as necessary.} 
\end{center} 
\end{figure*}

\subsection{Model atmosphere parameters}

Model atmosphere parameters -- effective temperature, surface gravity, microturbulence, and
metallicity --
are determined from the Fe\,{\sc i} and Fe\,{\sc ii} lines (Table 4). The selection of 28
Fe\,{\sc i} lines range in lower excitation potential (LEP) from 1.0 to 4.6 eV with equivalent
widths (EWs) of up to 270 m\AA\ but only four lines are stronger than 200 m\AA. Eleven Fe\,{\sc
ii} lines have LEPs from 2.8 to 6.7 eV with  EWs of up to 239 m\AA. Observed
and computed line profiles of the chosen Fe\,{\sc i} and Fe\,{\sc ii} lines are shown in Figures
4 and 5.

First, the temperature is estimated by requiring that derived abundances are independent of the
lower excitation potential (LEP). The microturbulence $\xi$ is derived by requiring that the derived abundances
are independent of line strengths. For our sample of Fe\.{\sc i} lines, these two conditions
are imposed simultaneously. We have used Fe\,{\sc ii} lines for measuring $\xi$ microturbulence
since appreciable departures
from  {\sc LTE} may occur for Fe\,{\sc I} lines (Boyarchuk et al. 1985; Th\'{e}venin \& Idiart 1999).
In this calculation, $\xi$ is assumed to be depth independent and isotropic. A check on the
microturbulence is provided by the lines of other species using the dispersion in the
abundances over a range in the $\xi$ for a given model -- see Fig.6. The dispersion $\sigma$ for Fe\,{\sc I}, Fe\,{\sc ii} and Ca\,{\sc i} lines are
computed for two different effective temperature values: \Teff = 7350 \kelvin\,(e.g. black
curve for Fe and Ca) and 5800 \kelvin\, (e.g. red curve for Fe and Ca) but since $\xi$ far
exceeds the thermal velocities, the result is essentially independent of $\Teff$. We adopt 6.6 $\pm$ 1.6 km s$^{\rm
-1}$. An estimate of the gravity is provided by the familiar requirement that Fe\,{\sc i} and
Fe\,{\sc ii} lines provide the same Fe abundance to maintain the ionization equilibrium (Fig.
7). Since these atmospheric parameter are interdependent, several iterations have been
performed to determine a suitable model from the model atmosphere grid.

The atmospheric parameters obtained are $\Teff$ = 7350 \kelvin\ , \logg = 0.6, [Fe/H]=0.4, and $\xi$ =
6.6 km s$^{\rm -1}$. The corresponding iron abundance is  $\log\epsilon(Fe)=7.93$ or [Fe/H]$\approx$0.4 for
the solar abundance of $\log\epsilon(Fe)=7.50$ (Asplund et al. 2009). The uncertainty in the derived surface temperature is provided by the error in the slope of the
relation between the Fe\,{\sc i} abundance and LEPs of the lines. A perceptible change of slope occurs
for a variation of $\pm$200 \kelvin\, in the adopted model (see Fig.7 top panel). In a similar way, 1-$\sigma$ abundance
difference [X/H] between neutral and ionized lines of Fe corresponds to a change of 0.3 dex in \logg.
The abundances of other elements were derived (Table 6).

After having determined microturbulence, we attribute the residual broadening needed to fit the
observed line profiles to macroturbulence. For determination of $\xi_{\rm mac}$, we used a sample of nine
relatively low excitation lines of Fe\,{\sc i} insensitive to collisional broadening located at 5300 \AA\,,
6000 \AA\,, and 6100 \AA\,. For each line,
we changed $\xi_{\rm mac}$ until the observed half width agreed with the calculated
one.\footnote{The {\sc MOOG} code computes a radial-tangential macroturbulence profile based on
the work of Gray 1992, in "The Obs. \& Anal. of Stell. Phot", p. 409.} The mean
$\xi_{\rm mac}$ resulting from this procedure after correcting for the microturbulence (6.6 km s$^{\rm
-1}$) and the instrumental width (5 km s$^{\rm -1}$), is $\xi_{\rm mac}$= 22.0$\pm$1.7 km s$^{\rm
-1}$. Projected rotational velocity is assumed to be negligible.  

Following analysis of the 2009 May 9 McDonald spectrum, we determined model atmosphere parameters from all
McDonald and SAO spectra -- see Table 7. There is no secular trend in any of the atmospheric parameters. So, one
can conclude that the line profile variations are not reflected in the atmospheric
parameters. For instance, the
2010-2011 episode of H$\alpha$ broadening presented in Figure 2 is not seen in the model parameters from the
McDonald and SAO spectra.

\begin{figure*} \begin{center}
\includegraphics[width=180mm,height=125mm,angle=0]{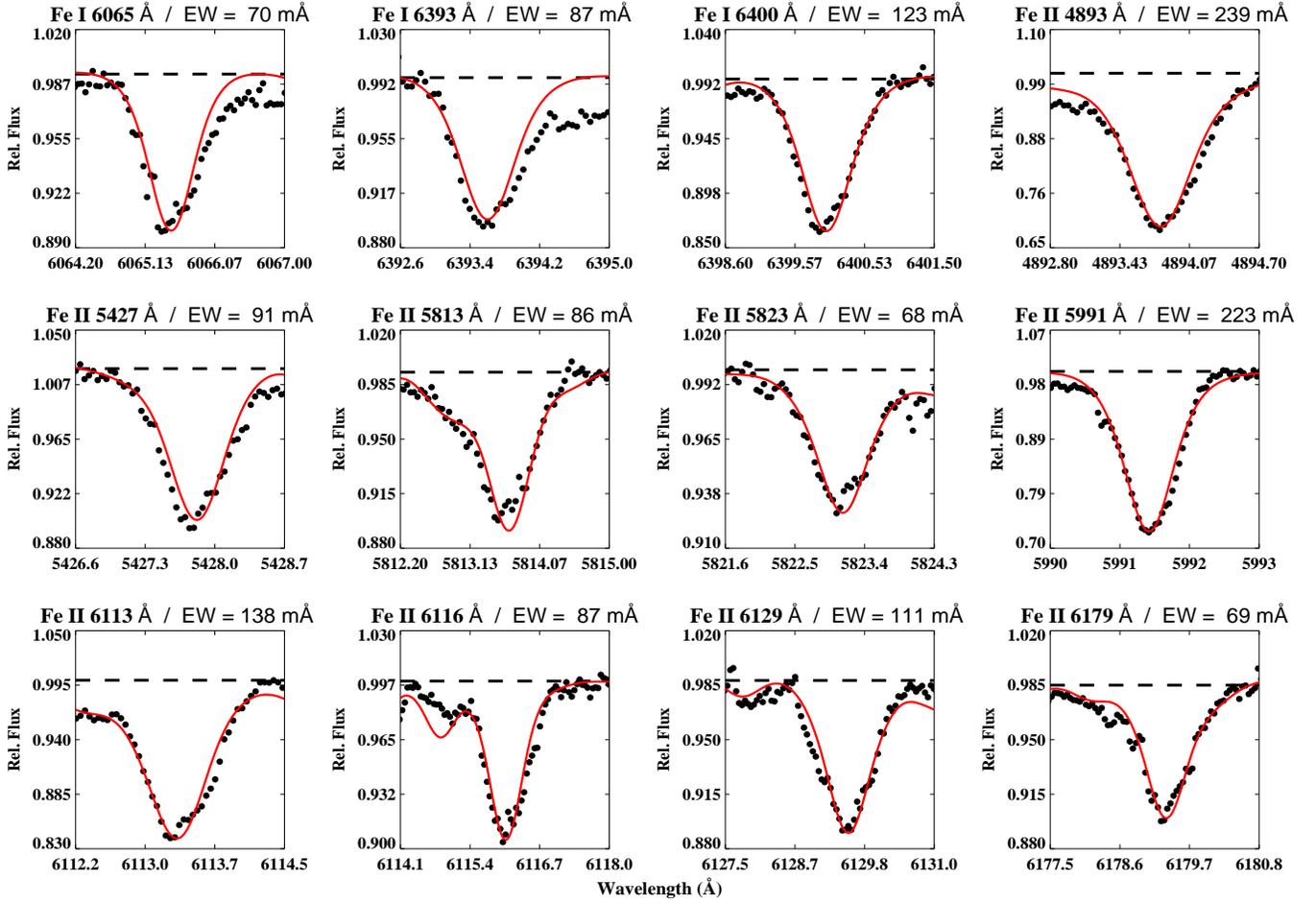} 
\caption{As in Fig.4 but Fe\,{\sc ii} lines are included (see caption to Figure 5).} 
\end{center} 
\end{figure*}

\subsection{Balmer and Paschen lines}

Additional information about the atmosphere is provided by the hydrogen Balmer and
Pashen lines. At the parameters of HD\,179821, the line profiles are sensitive to 
temperature and gravity. The variable H$\alpha$ profiles (Figure 2) are clearly
highly distorted modifications of the photospheric profile. Thus, we seek to constrain the
atmospheric parameters using the Balmer lines H$\beta$, H$\gamma$  and H$\delta$ and
the Paschen lines P$\delta$ at 10049\AA\ and P11 at 8863\AA. These profiles appear immune to the extreme variations exhibited by H$\alpha$ -- see Figure
8 (top right) for a selection of spectra around the H$\beta$ line.

\begin{figure}
\begin{center}
\includegraphics[width=84mm,height=70mm,angle=0]{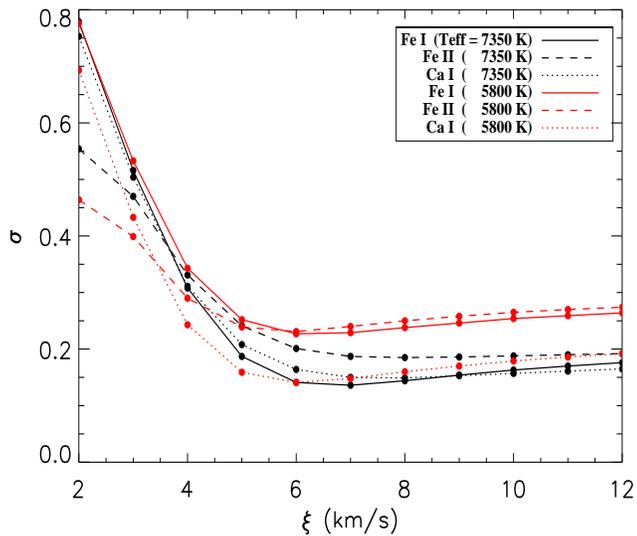}
\caption{The standard deviation of the Fe and Ca abundances from the suite
of Fe\,{\sc i}, Fe\,{\sc ii}, and Ca\,{\sc i} lines as a function of the microturbulence $\xi$.}
\end{center}
\end{figure}

In the top left panel of Figure 8, we superimpose on the H$\beta$ profile for the McDonald
2009 May 9 spectrum theoretical profiles for [Fe/H]=+0.4 models for T$_{\rm eff} = 7350$ K and $\log g = 0.4$ and
$0.6$  and  T$_{\rm eff} = 7550$ K and $\log g = 0.4$. The observed red wing is shallower
than predicted suggesting that emission is affecting this wing. Predicted profiles were
computed with {\sc SYNTHE} and convolved with a Gaussian profile in {\sc DIPSO} to simulate the
instrumental broadening. For all predicted profiles blending lines were computed with
abundances scaled to the metallicity [Fe/H] $= +0.4$. These are LTE profiles but non-LTE
profiles for similar supergiant atmospheres suggest that corrections for non-LTE effects
affect the core of the line and are likely to be very small if collisions with hydrogen atoms
are included in the statistical equilibrium calculations for the H atom (Barklem 2007;
Barklem 2016, private communication). The best fit to H$\beta$'s blue wing is found for T$_{\rm
eff} = 7550\pm200$ K and $\log g = 0.4\pm0.3$. These parameters are in fair agreement with the
spectroscopic values from iron lines. Examination of a family of predicted profiles shows
that H$\beta$ can be fit by profiles along a locus in the (T$_{\rm eff}, \log g)$ plane from
(7350, 0.3) to (7750, 0.6) through (7550,0.4). H$\gamma$ and H$\delta$ as well as Paschen lines
are well fitted also by the (7550,0.4) model and other models along the locus of best-fits to
the H$\beta$ line. Emission in the red wings is absent for H$\beta$ and H$\gamma$ but,
perhaps, weakly present for the Paschen lines (Figure 8).

Previous  fits to the Balmer line H$\delta$ gave lower temperatures which were considered to
be at odds with the spectroscopic determination: Reddy \& Hrivnak (1999) gave (6750,0.5) and
Kipper (2008) obtained (6000,2.0).  These estimates straddle the locus found from our fit to
the H$\beta$ line.

\begin{figure}
\begin{center}
\includegraphics[width=84mm,height=70mm,angle=0]{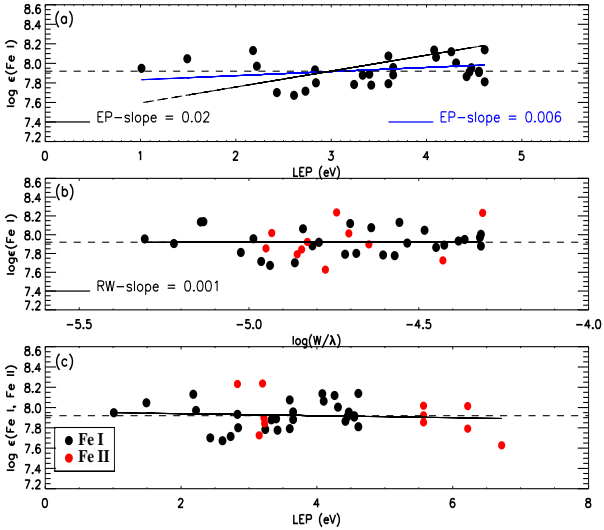}
\caption{Determination of atmospheric parameters $\Teff$, $\xi$, and $\logg$ using abundance (log
$\epsilon$(Fe)) as a function of both lower level excitation potential (LEP) and reduced equivalent
width (R.W., log(W/$\lambda$)). In all the panels, the solid line is the least-square fit to the data.
In (a), the solid black line is for $\Teff=$7350 $\kelvin$ and the blue line is for
$\Teff=$7550 $\kelvin$ and in (b), log $\epsilon$(Fe) vs.
log $W/\lambda$ suggests $\xi$=6.6 km s$^{\rm -1}$. In (c), log $\epsilon$(Fe\,{\sc i})=log
$\epsilon$(Fe\,{\sc ii}) is achieved for $\logg$=0.6.}
\end{center}
\end{figure}

\subsection{The Chemical Composition}

In Table 6, we presents a summary of the elemental abundances based on the LTE based model
parameters. In Table 6, $\log\epsilon$ is the logarithm of the abundances. The $\sigma_{\rm line}$ is
the $1\sigma$ line-to-line scatter in the abundances. The $[X/H]$ is the logarithmic abundance ratio
with hydrogen relative to the corresponding solar value, and $[X/Fe]$ is the logarithmic abundance
with respect to the Fe\,{\sc i} abundance. Estimated formal errors for the abundances arising from
uncertainties of the atmospheric parameters \Teff, \logg, and $\xi$ are summarized in Table 8 for
changes with respect to the model of $+200$ \kelvin, $+0.3$ cgs units, and $\pm1.6$ km s$^{\rm -1}$.
From the uncertainties listed in Table 8, we find the total absolute uncertainty ($\sigma_{\rm abs}$)
to be range from 0.06 for O\,{\sc i} to 0.36 for La\,{\sc ii} by taking the square root of the sum of
the square of individual errors (for each species) associated with uncertainties in temperature,
gravity, and microturbulent velocity (see also column 4 in Table 6). The $\sigma_{\rm [X/H]}$
presents the error in [X/H] and is the square root of the sum of the quadrature of the errors in
stellar and solar logarithmic abundances. The $\sigma_{\rm [X/Fe]}$ is the error in [X/Fe] and the
square root of the sum of the quadratures of the errors in $[X/H]$ and $[Fe/H]$.     

\begin{figure*} \begin{center}
\includegraphics[width=180mm,height=117mm,angle=0]{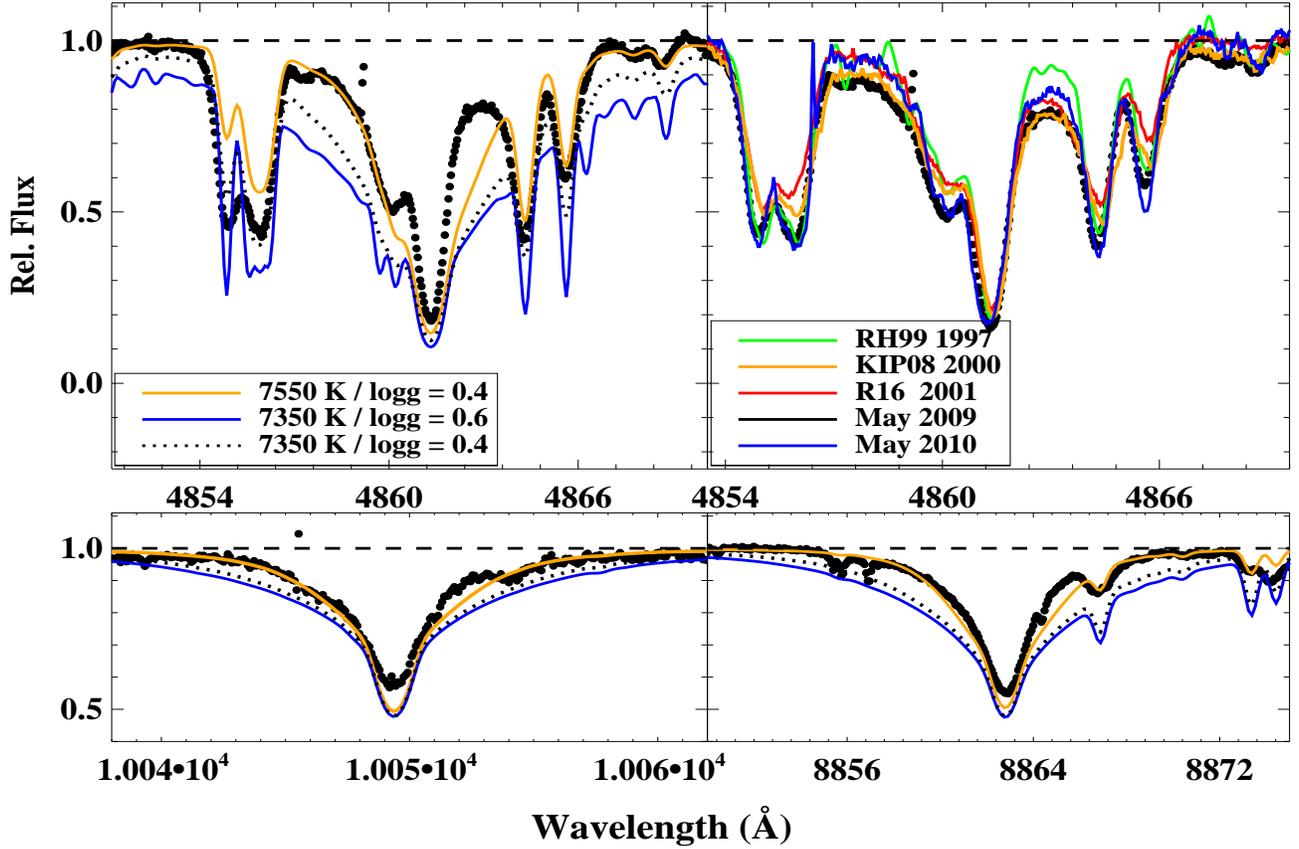} 
\caption{The left panel shows the observed (black dotted) and the model line profiles for H$\beta$.
The theoretical profiles have been generated for a surface gravity \logg = 0.4 and 0.6 dex,
\Teff = 7350 \kelvin\, and 7550 \kelvin, and [Fe/H]=+0.4. The right panel shows the observed H$\beta$ profiles for the
1997 October McDonald spectrum (Reddy 2016, private communication), the 2000 Oct ELODIE spectrum
(Kipper 2008), the 2001 August McDonald spectrum (R16; Reddy 2016, private communication), and the
2009 and  2010 May McDonald spectra. The lower panels shows the observed and the model line profiles
for the Paschen lines P$\delta$ at 10049\AA\ and P11 at 8863\AA.} 
\end{center} 
\end{figure*}

Inspection of Table 6 reveals several interesting aspects of HD
179821's composition. The mean metallicity is [Fe/H] $= +0.33$ from all spectra - see Table 7. Luck (2014) assembled and analyzed a sample of 451 F, G, and K  stars of luminosity
classes I and IIa of which the vast majority are evolved intermediate mass stars and not post-AGB
stars. A handful had [Fe/H] of 0.4 or greater but in conformity with expectation these stars were at
Galactocentric distances inside the Solar circle. At HD 179821's Galactic longitude of 36$^\circ$,
the line of sight at closest approach to the Galactic centre is at 5 kpc for a distance from the Sun
of slightly less than 7 kpc. Given that there is an abundance gradient with higher metallicities
toward the Galactic centre, a positive [Fe/H] is not unexpected. Luck and Lambert (2011) from Cepheid
variables obtained the slope $d[Fe/H]/dR_G = -0.061$ dex kpc$^{-1}$ which gives  [Fe/H] $\simeq 0.2$
for a star at 5 kpc. In Luck's sample, almost all supergiants inside Galactocentric distances of 7
kpc have [Fe/H] of 0.3 or greater, and these values are in excess of those expected from the gradient
and [Fe/H] = 0 at the Sun's Galactocentric distance of 8.5 kpc.  Thus, at [Fe/H] = $+0.4$, HD
179821 may not be an unusual supergiant.

What may be thought unusual is that the metallicity [Fe/H] $= +0.4$
is significantly higher than all previous spectroscopic
determinations. Considering that such determinations have used spectra
of comparable quality to our and methods of similar approach, it is
necessary to reexamine published metallicity determinations which we
do in the next section.

\subsection{The metallicity [Fe/H]}

Determinations of HD 179821's composition from high-resolution optical
spectra  have been  reported by Za\v{c}s  et al. (1996), Reddy \& Hrivnak
(1999), Kipper (2008) and Luck (2014). Behind these (and our) analyses
is a common framework involving the combination of plane-parallel model
atmospheres in LTE and hydrostatic equilibrium with a line analysis
programme also based on the assumption of LTE. Analyses differ by dates of the observations, the  wavelength coverage of the spectra, the selection
of lines and $gf$-values. 

HD 179821 has presented two
different faces to the quantitative spectroscopist: it is either metal-poor
with [Fe/H] $\simeq -0.1$ to $-0.4$ with atmospheric parameters
T$_{\rm eff} \simeq 6800$ \kelvin\, and $\logg \simeq  0.5$ to $1.3$
(Za\v{c}s et al. 1996; Reddy \& Hrivnak 1999;
Kipper 2008) or it is metal-rich with [Fe/H] $\simeq +0.4$ with
atmospheric parameters T$_{\rm eff} \simeq 7300$ \kelvin\, with $\logg \simeq
0.6$ to 1.0 (Luck 2014; this paper)\footnote{Th\'{e}venin,
Parthasarthy, Jasniewicz (2000)'s analysis of a low-resolution
spectrum ($R \sim 8000$) gave [Fe/H] = $-0.5$ and T$_{\rm eff} = 5660$K
and $\log g = -1.0$.}.

Za\v{c}s et al. analysed a spectrum from 1992 August 17 obtained at the {\sc SAO}. Abundances
were given relative to those of the G8IIIab giant $\epsilon$ Vir. With parameters $T_{\rm eff} =
5130$K, $\logg = 2.5$ and $\xi = 2.0$ km s$^{-1}$ and [Fe/H] = 0, HD\,179821 was found to be
slightly Fe-poor with [Fe/H] $= -0.1$. However, chosen parameters for the star do not satisfy
ionization equilibrium for either Fe or Cr: the reported mean Fe abundance was 7.31 from
Fe\,{\sc i} and 7.69 from Fe\,{\sc ii} lines. In light of this discrepancy, we reanalysed
Za\v{c}s et al.'s line list, and obtained the model parameters: \Teff = 7300 \kelvin\,, \logg =
0.60, [Fe/H] $= +0.40$, and $\xi = 6.50$ km s$^{\rm -1}$. These parameters are compatible with
our results from McDonald and SAO spectra (Table 7).

The McDonald spectra of 1997 October 15-16 taken by Reddy \& Hrivnak
were analyzed using our line list. Parameters: T$_{\rm eff} = 7150$ \kelvin\,,
$\logg = 0.5$, a microturbulence $\xi = 9.7$ km s$^{-1}$ and [Fe/H] $= +0.20$
were obtained. This temperature is 400 K hotter
and the metallicity 0.3 dex greater than reported by Reddy \& Hrivnak.
Examination of the Fe lines chosen by
Reddy \& Hrivnak found that 9 of 23 Fe\,{\sc i} and 3 of 9 Fe\,{\sc ii}
did not meet our criteria of a measurable and unblended line. We
suggest that the sub-solar Fe abundance reported by Reddy \&
Hrivnak  likely  arises from an imperfect list of Fe lines.

Kipper (2008) analyzed spectra acquired in 2000 September and October and
stored in  the ELODIE archive (Moultaka et al. 2004). He compiled an extensive
line list which was analysed using the model selected by Reddy \& Hrivnak
(1999): T$_{\rm eff} = 6750$ \kelvin\,, $\logg = 0.5$ but with a slightly higher
microturbulence $\xi = 6.6$ km s$^{-1}$. He reported a mean Fe
abundance of $\log\epsilon$(Fe) = 7.00$\pm0.20$ from 36 Fe\,{\sc i}
lines and 11 Fe\,{\sc ii} lines but ionization equilibrium was not achieved:
the Fe abundance was 6.90 from Fe\,{\sc i} and 7.16 from Fe\,{\sc ii} lines.
Relative to Asplund et al.'s solar Fe abundance, Kipper's mean value
corresponds to [Fe/H] $= -0.5$. 

On subjecting Kipper's line list to our analysis procedures, we found
a model providing excitation and ionization equilibrium, namely
T$_{\rm eff} = 7450$ \kelvin\,, $\logg = 0.60$, $\xi = 8.6$ km s$^{-1}$ with
Fe abundances of 7.88$\pm0.17$ and 7.87$\pm0.22$ from the Fe\,{\sc i}
and Fe\,{\sc ii} lines, respectively, i.e., [Fe/H] $= +0.47$. These
parameters and Fe abundance are thoroughly compatible with our
results from McDonald and SAO spectra (Table 7).

When we ran Kipper's Fe line list through the Reddy \& Hrivnak model,
we found the same lack of ionization equilibrium but higher Fe
abundances than reported by Kipper. On examination of the complete line list, we found
that our abundances were $0.44\pm0.05$ dex greater across the array of
elements; this surprising  difference comes from using the same line list as
Kipper in every respect and a Kurucz model with Kipper's adopted parameters. This 0.4 dex difference is not understood. (In order to satisfy a
curiosity, we extracted the 2000 ELODIE spectra used by Kipper from the archive and
measured a set of clean lines and compared equivalent widths (EWs).
Our measurements are in fair agreement with Kipper's: 
EW(K) = 1.3($\pm0.07$)EW(Us) + 1.6($\pm6.9$). Such EW differences have nothing
to do with the 0.44 dex difference above.) 

Our final comparison is with Luck (2014) who analyzed ELODIE spectra\footnote{EWs averaged over seven ELODIE spectra, 
including the same spectrum as that analyzed by Kipper (2008), with the two highest S/N (153 and 180) having weight 2 and
the others weight 1.(Luck 2016, private communication).} and one taken with the High-resolution spectrograph at 
Hobby-Eberly Telescope (HET) at the McDonald Observatory (Tull 1998). The HET spectrum is from 2010 July 21.  
Luck analyzed both the ELODIE and the HET spectra using models
from two grids: the Kurucz ATLAS and the Uppsala MARCS grids.  
The selected ATLAS model had T$_{\rm eff} = 6997$K, $\log g = 0.62$
and a $\xi =4.76$ km s$^{-1}$. The chosen MARCS model was similar with
T$_{\rm eff} = 7107$K, $\log g = 1.0$ and $\xi = 4.74$ km s$^{-1}$.
The metallicity [Fe/H] was +0.5 from the ELODIE spectrum with just a 0.03 dex 
difference between the values from the ATLAS and MARCS models. A slightly
lower [Fe/H] value  of $+0.35$ was obtained from the HET spectrum, again with
a 0.03 dex difference between the two models. Thus, Luck's analyses confirm
our results for the atmospheric parameters and metallicity [Fe/H]. 

In short, reanalysis of published spectroscopic analyses show that there is general
agreement that HD\,179821 is metal-rich [Fe/H] $\simeq +0.4$. In the next section, we discuss
relative abundances [X/Fe].

 \begin{table}
\begin{center}
    \caption[]{Comparison of abundances of the observed species for HD\,179821.}
       \label{}
   $$
       \begin{array}{l@{}||cc||cc|cc}
          \hline
           &\multicolumn{2}{c||}{\mbox{This\,work}} & \multicolumn{4}{c}{Luck (2014)}\\ 
\cline{2-7}
 $Element$ &$[X/Fe]$ & \sigma_{\rm abs} & $[X/Fe]$^{\rm 1}& \sigma_{\rm 1} &$[X/Fe]$^{\rm 2} & \sigma_{\rm 2}\\	 
\cline{2-7}
           &$(dex)$ &$(dex)$ &$(dex)$& $(dex)$ & $(dex)$ & $(dex)$	 \\ 
\hline
 C$\,{\sc i}$   &$-$0.19 &0.18&$-$0.58 & ... &$-$0.28 &0.10 \\
 N$\,{\sc i}$   &   0.87 &0.25&1.78    & ... &   0.76 &0.17 \\ 
 O$\,{\sc i}$   &$-$0.25 &0.10&$-$0.38 & ... &   0.00 &0.11 \\
 Na$\,{\sc i}$  &   0.98 &0.31&1.01    & 0.24&   0.48 &0.12 \\
 Mg$\,{\sc i}$  &   0.19 &0.32&0.17    & 0.46&   0.17 &0.12 \\
 Al$\,{\sc i}$  &$-$0.17 &0.20&0.16    & 0.00&   0.15 &0.16 \\
 Si$\,{\sc i}$  &   0.50 &0.23&0.26    & 0.13&   0.20 &0.03 \\
 S$\,{\sc i}$   &   0.48 &0.21&0.50    & 0.54&   0.11 &0.09 \\
 Ca$\,{\sc i}$  &   0.16 &0.33&$-$0.18 & 0.81&   0.03 &0.05 \\
 Sc$\,{\sc ii}$ &   0.26 &0.17&0.29    & 0.18&   0.43 &0.18 \\
 Ti$\,{\sc ii}$ &   0.26 &0.21&0.05    & 0.24&   0.30 &0.10 \\
 Cr$\,{\sc i}$  &$-$0.03 &0.29&0.68    & 0.63&   0.23 &0.08 \\
 Cr$\,{\sc ii}$ &$-$0.05 &0.15&0.17    & 0.00&   0.23 &0.08 \\
 Mn$\,{\sc i}$  &$-$0.34 &0.27&0.13    & 0.35&   0.07 &0.10 \\
 Fe$\,{\sc i}$  &   0.00 &0.26&$-$0.01 & 0.24&   0.07 &0.05 \\
 Fe$\,{\sc ii}$ &$-$0.10 &0.09&0.08    & 0.23&   0.07 &0.05 \\
 Ni$\,{\sc i} $ &   0.28 &0.25&0.26    & 0.42&   0.05 &0.07 \\
 Zn$\,{\sc i}$  &$-$0.11 &0.26&0.09    & 0.55&$-$0.21 &0.09 \\
 Y $\,{\sc ii}$ &$-$0.04 &0.21&0.26    & 0.78&   0.51 &0.39 \\
 Zr$\,{\sc ii}$ &   0.29 &0.15&$-$0.04 & 0.34&   0.24 &0.18 \\
 Ba$\,{\sc ii}$ &$-$0.01 &0.29&$-$0.31 & 0.00&   0.12 &0.17 \\
 La$\,{\sc ii}$ &   0.01 &0.36&$-$0.17 & 0.45&   0.07 &0.12 \\
 Nd$\,{\sc ii}$ &   0.18 &0.27&$-$0.01 & 0.38&   0.03 &0.16 \\
 Eu$\,{\sc ii}$ &   0.69 &0.20&0.48    & 0.38&   .... &...  \\			    
\hline
       \end{array}
   $$
\begin{list}{}{}
\item {\bf ($^{\rm 1}$):} Mean abundances for HD\,179821 from 2010 Hobby-Eberly Telescope (HET)
spectrum; Luck (2014).
\item {\bf ($^{\rm 2}$):} Mean abundances from Luck (2014) for a subsample of 8 stars analysed with MARCS grids
and with [Fe/H]
$\ge$ 0.0 dex and $\sigma_{\rm 2}$ is star-to-star scatter in [X/Fe] values.
\end{list}
\end{center}
 \end{table}

\subsection{Relative abundances [X/Fe]}

In principle, HD 179821's chemical composition may offer insights into the status of the
star: a massive post-main sequence star or a lower mass post-AGB star. Obviously, such
insights are compromised by uncertain and erroneous abundances. In order to minimize
compromises, we pursue a multi-part discussion. We compare our abundances (Table 6) with
those from Luck (2014) who, as noted above, undertook a large survey of F-G supergiants and
included HD\,179821.  
In Table 9, we compare our [X/Fe] with the average provided by Luck from the HET spectrum
and the MARCS model. Given the estimated $\sigma$s the comparison suggests fair
agreement. A notable feature is the agreement that Na (relative to
Fe) is highly overabundant, a feature noted by all previous analyses of HD\,179821. 
In the case of Cr where Luck's [Cr/Fe] is 0.7 dex greater than ours and his [Cr/Fe] from
Cr\,{\sc i} and Cr\,{\sc ii} lines differ by 0.6 dex, it is possible that his limited
selection of Cr\,{\sc i} lines contains blended lines.

Luck's sample of supergiants is dominated by evolved massive stars and many have higher
surface gravities than HD 179821. Yet, it is instructive to compare abundances obtained for
HD\,179821 with selected samples drawn from Luck's large survey.  The sample of eight stars
represented in Table 9 have T$_{\rm eff}$ from 6600-7200 K and $\logg$ from 1.1- 2.0. There
is general agreement with our results for HD\,179821, notably for C, N, and O but
interesting disagreements for Na, S and Y.  The high Y abundance in the sample is
likely a matter of line selection given that Zr does not share the apparent overabundance,

HD\,179821's [X/Fe] may be judged against expected values for a star with [Fe/H] $\sim +0.4$.
Abundance analyses of local dwarfs and giants show that for Na to Zn, relative
abundances [X/Fe] do not differ greatly from zero, even at [Fe/H] $\simeq +0.4$, the
metallicity of HD\,179821 -- see, for example,  Bensby et al. 2014) for dwarfs and Luck \&
Heiter (2007, their Table 10) for giants. Although HD\,179821's metallicity places it at or
even beyond the high metallicity limit of these samples, [X/Fe] $\sim 0.0$ might be expected.
With respect to this baseline, inspection of our and Luck's results in Tables 6 and 9 shows
the one outstanding anomaly is for Na with [Na/Fe] $\simeq 1.0$. Possible additional
anomalies include [Si/Fe], [S/Fe], [Sc/Fe], and [Ni/Fe] with [X/Fe] $\sim +0.3$ to $+0.5$. Table 6 shows
that HD\,179821 is not enriched in $s$-process products. Eu, an r-process product; appears
overabundant relative to the anticipated [X/Fe] $\sim 0.0$.  

The assumption of LTE was adopted in the construction of the model atmosphere and in the
analysis of the absorption lines. Given the low particle densities in the atmosphere, one
should be concerned about the effects of departures from LTE on the atmospheric structure and
the formation of the lines. We are unaware of supergiant atmosheres constructed in non-LTE. 
There are some calculations of line formation in relevant atmospheres which suggest
approximate corrections to [X/H] and [X/Fe] for HD\,179821. Fortunately, some results are
available for the interesting light elements C, N and O. Venn (1995) in her analyses of A and
F supergiants computed non-LTE corrections for selected lines of C\,{\sc i} and N\,{\sc i}
and her stellar sample included four stars with T$_{\rm eff}$ between 7400-7600 \kelvin\, and
surface gravities $\log g$  from 1.1-1.6 with approximately solar metallicities. Considering
our chosen lines and assuming that Venn's quartet are representative of HD\,179821, Venn's
calculations imply that the non-LTE abundance is about 0.3 dex for both C and N smaller than
the tabulated LTE abundances. Takada \& Takeda-Hidai (1998) provide non-LTE predictions for
the 6157 \AA\, lines in A-F supergiants, At the atmospheric parameters of HD\,179821, their
calculations show that the LTE O abundance should be reduced by 0.1 to 0.2 dex. 

For abundances of species from Na to Zn, the outstanding abundance anomaly is held by Na with a 1.0 dex enrichment
which may be an indicator for a high luminosity status of the star, hence operation of Ne-Na
cycle (Denissenkov \& Ivanov 1987; Denissenkov 2005). Lind et al.'s (2011) extensive non-LTE
calculations for Na\,{\sc i} lines did not cover the atmospheric parameters of HD\,179821 as
the most extreme supergiant model was a relatively cool T$_{\rm eff} = 5500$K  at $\logg =
1.0$. By extrapolation, it would seem that the non-LTE correction to HD\,179821's Na
abundance is small. This suspicion is confirmed by pioneering calculations by Boyarchuk,
Denisenkov, \& Hubeny (1988a) and Boyarchuk et al. (1988b, c): Korotin (2016, private
communication) predicts $\Delta([Na/H]) = $ -0.14 dex\footnote{$\Delta([Na/H])=[Na/H]_{\rm
NLTE}-[Na/H]_{\rm LTE}$}. Venn's (1995) estimates for a few Mg\,{\sc i} lines in late-F and
early-F supergiants suggest only a 0.1 dex reduction of the LTE abundances but her and our
selections of Mg\,{\sc i} lines show little overlap. Extensive calculations of Mg\,{\sc i}
line formation in cool stellar atmospheres have been reported by Osorio \& Barklem (2016) for
many Mg\,{\sc i} lines. Inspection of their predictions for four strong lines in Table 3 show
that the non-LTE Mg abundances for a HD\,179821-like model atmosphere\footnote{The closest 1D
model atmosphere is the (\Teff, \logg, [Fe/H]) = (7250, 1.5, 0.5) model.} are about +0.2 dex
greater than the LTE value. The highest correction is needed for 5528 \AA\, Mg\,{\sc i} line
with +0.2 dex and the corrections for 4057, 4167, and 4702 Mg\,{\sc i} lines are about +0.1
dex. These corrections are less than the uncertainty arising from the microturbulence.
Non-LTE calculations for Zn\,{\sc i} lines (Takeda et al. 2005) show a very small non-LTE
abundance correction. Non-LTE effects for prominent Ba\,{\sc ii} lines in cool stars have
been calculated by Korotin et al. (2015). The examined grid of stellar atmospheres extends up
to 6500 \kelvin\, and down to $\log g = 0$ and to stars as Fe-rich as [Fe/H] = +0.5. For the model
parameters reported in Section 4.2, the non-LTE Ba abundance from the 5853 \AA\ Ba\,{\sc ii}
line for HD\,179821 is +0.3 dex greater than the LTE abundance listed in Table 6 (Korotin
2016, private communication). In interpreting [X/Fe], one obviously must consider the non-LTE
corrections to the Fe abundance. In the case of Fe, a leading non-LTE effect is the
over-ionization of neutral iron atoms leading to an underestimate by the LTE analysis of the
Fe abundance from Fe\,{\sc i} lines, Lind et al.'s (2012) extensive calculations of non-LTE
effects across the Fe\,{\sc i} spectrum  did not unfortunately extend to an atmosphere
representative of HD\,179821. At T$_{\rm eff} > 7000$ \kelvin, models considered had $\log g \geq
3$ and [Fe/H]  $\leq +0.25$.  A necessarily crude extrapolation suggests the LTE Fe abundance
is underestimated by less than 0.1 dex. Species similar to Fe  (e.g., Ni) will presumably be
affected similarly and [X/Fe] will require an even smaller correction for non-LTE effects. 
We have not considered the role these effects play in determining the atmospheric parameters
from Fe\,{\sc i} and Fe\,{\sc ii} lines (Lind et al. 2012). 

%The non-LTE correction to Mg abundance for HD\,179821 by Korotin (2016, private
%communication) is +0.38 dex (i.e. $\Delta([Mg/H]) = $ -0.38 dex).
Non-LTE estimates for C, N, O, Na, Mg, Fe, Zn, and Ba appear to exhaust what has been
published for warm supergiants such as HD\,179821. In making this
claim we have consulted the valuable review by Bergemann \& Nordlander (2014)
with its table of non-LTE studies of late-type stars. Many of the
cited papers are concerned primarily or exclusively with the solar spectrum and several of the
very recent papers focus on just five stars (Procyon, Arcturus, HD\,84937, HD\,140283
and HD\,122563) from which extrapolation to a supergiant like HD\,179821 is most uncertain.

In the quest for more accurate abundance determinations for HD\,179821, incorporation of
non-LTE effects in the interpretation of the absorption lines and even in the construction of
a model atmosphere may not provide the biggest leap to the end. The star's atmosphere
violates the standard assumption of uniform plane-parallel (or spherical) layers in
hydrostatic equilibrium. A supersonic macroturbulence of 22 km s$^{-1}$ starkly contradicts
this assumption and challenges theoretical stellar astrophysicists to build more inclusive
model atmospheres. The important STAGGER grid (Magic et al. 2013) does not intrude into the
domain belonging to HD\,179821; models at 7500 \kelvin\, refer to main sequence stars and the lowest
gravity models ($\log g = 1.5$) are no hotter than 4500 \kelvin. Realistic atmospheres of 
supergiants will have many applications including the extension of determination of analyses
of supergiants to several galaxies beyond the Galaxy. 

\section{Concluding Remarks}

There is no question but that HD\,179821 is a luminous star. Possible identifications of
the star include two possibilities: (i) a massive star evolving from the main sequence at
approximately constant luminosity to the red supergiant phase or on a post-red supergiant loop
back to the blue or (ii) a lower mass star evolving at roughly constant luminosity from the AGB to
the tip of the white dwarf cooling track.  

The mass-luminosity relations for these alternative identifications overlap
for a range of masses. Above a certain
luminosity, the more likely identification is a high mass star. This critical luminosity
is set by the maximum mass -- the Chandrasekhar mass -- of a post-AGB star
which may become a white dwarf. At the Chandrasekhar mass, the luminosity of
a post-AGB star is M$_{bol} \simeq $-7.1 or $\log L/\Lsolar \simeq $4.7
(Wood, Bessell, \& Fox 1983). Therefore, if it can be shown that HD 179821's
luminosity exceeds the latter limit by a clear margin, one may identify the
star as a massive star.

Estimations of HD\,179821's absolute magnitude from an apparent  magnitude are
fraught with uncertainty owing to an uncertain correction for
interstellar extinction with the possibility of an additional correction
for circumstellar extinction. As previous investigators of the star
have appreciated (Reddy \& Hrivnak 1999; Kipper 2008; Oudmaijer et al. 2009), the absolute magnitude of HD 179821 may be
estimated from the equivalent width (EW) of the O\,{\sc i} triplet
at 7770-7774 \AA. The $M_V$ -- $EW$ calibration for
warm supergiants comes from Kovtyukh, Gorlova \& Belik (2012) who
assembled EW measurements for supergiants with known luminosities. Our
measurement of the oxygen triplet's EW is 2.7 \AA. Kovtyukh et al.'s
calibration has few stars with such strong EW but  mild extrapolation of
the calibration gives $M_V \simeq $-8.9 or $\log L/\Lsolar\simeq $5.5.
This absolute magnitude is nearly two magnitudes brighter than the maximum
for a post-AGB star and slightly fainter than
the Humphreys-Davidson (Humphreys \& Davidson 1979) limit of M$_{\rm bol} \simeq $-9.5 for the most
luminous warm Galactic supergiants such as $\rho$ Cas and HR 8752.

Absolute luminosities for post-AGB stars are poorly known, in general. Arellano Ferro, Giridhar,
Arellano (2003) incorporated five post-AGB stars into their calibration of $M_V$ -- $EW$
relation based primarily on normal supergiants. The two post-AGB star with estimated
luminosities close to the luminosity limit for post-AGB stars had $EW$s of the triplet of 2.0
and 1.7 \AA, values similar to $EW$s of normal supergiants of the same luminosity. Thus, it
appears that normal supergiants and post-AGB stars of type F-G satisfy similar $M_V$ -- $EW$
relations for the oxygen triplet. 

The effective temperature and surface gravity provide a check on the
conclusion that HD 179821 is a massive star. By combining the relations
$L \propto R^2T_{\rm eff}^4$ and $g \propto M/R^2$, one
obtains

\begin{equation}
\hskip 1.5 cm log L/\Lsolar = log M/\Msolar  + 4log T_{\rm eff} - log g - 10.61
\end{equation}

On substituting T$_{\rm eff} = 7350$K, $\log g = 0.64$ and 
$\log L/\Lsolar = 5.5$, a mass $M = 19M_\odot$  is obtained. Stellar
evolutionary tracks (e.g., Iben 1985) imply that such a luminosity is
achieved at a higher mass --
say, 30$M_\odot$ -- but an adjustment of $\log g$ by just 0.2 dex provides
just such a mass\footnote{An independent estimate of the mass was also provided by Parsec isochrones (Bressan et
al. 2012) with the solar metal content of Z$_{\rm \odot} =$ 0.0152 and using a Z = 0.03
([Fe/H] = log(Z/Z$_{\rm \odot}$)) isochrone
(http://stev.oapd.inaf.it/cgi-bin/cmd-2.7). This isochrone, based on model atmosphere parameters reported
in Section 4.2, provided
a mass of 30 $\Msolar$ for HD\,179821.}.   

It remains to consider HD\,179821's composition in light of 
the proposed  identification as 
a massive star which has evolved beyond core H-burning on the main
sequence to a He-core burning  warm supergiant (and possibly beyond 
this phase!). A massive star observed as a warm supergiant may be expected to have
shuffled at a minimum its C and N abundances as CN-processed material
reached the surface by rotationally-induced mixing and the first dredge-up
with the latter a contributor if the star is now evolving to the blue
after a red supergiant phase. Present C/N/O abundances are taken from
our analysis summarized in Table 6 with the
approximate non-LTE corrections listed in Section 4.6, i.e., C = 8.5,
N= 9.2 and O = 8.9. On the assumption that initial abundances satisfied
the condition [X/Fe] = 0.0 and [Fe/H] = +0.4, C= 8.9, N = 8.5 and
O= 9.2 were the starting abundances; relative to these 
values N is clearly enriched, C and O are depleted. 
Assuming that the envelope has been
mixed with CN-processed material from the interior, the CN-cycle's
catalysts of C and N are required to be conserved. The initial 
(logarithmic) sum is 9.0 and the observed sum is 9.3 with conservation
satisfied within the measurement uncertainties and including the rough
corrections for non-LTE effects. For more severe mixing,
ON-cycled products may be involved and the conserved quantity is the sum of
the C, N and O abundances. In this case, the initial value is 9.4 and the
observed value is 9.3 which surely represents a case of fortuitous
agreement. (Note: the analysis does not consider enhancement of the
surface He abundance during evolution.) 

Sodium overabundances in F-G supergiants have been the subject of
several observational and theoretical investigations. As an
observational reference point for HD 179821, we take the survey 
by Andrievsky et al. (2002) of Na abundances in F-G Ib-II giants
which showed  [Na/Fe] increasing with decreasing $\logg$ reaching
[Na/Fe] $\simeq +0.3$ at $\logg = 1.0$. Non-LTE calculations show that
the departures from LTE on the observed Na\,{\sc i} lines at 6154 \AA\ and 
6160\AA\ are small ($\sim$ 0.1 dex)
for supergiants with $\Teff \sim 7000$ \kelvin.
The Na enrichment is attributed to operation of the H-burning Ne-Na chain
in which $^{22}$Ne is converted to $^{23}$Na.
Denissenkov (2005)  argued that observed levels of Na enrichment
required massive stars to undergo mixing between core and the
radiative envelope in their main sequence progenitor.  

Inspection of Andrievsky et al.'s list of stars showed that 
just one was as Fe-rich as HD\,179821. A reference sample of four stars
was selected with the conditions that $\Teff \geq 6500$ \kelvin\,  
and $\logg \le 1.0$. For this quartet, mean values are  [Fe/H] $= -0.24$ 
and [Na/Fe] $= +0.37$ with a small star-to-star scatter. If we assume that
the $^{22}$Ne abundance scales with [Fe/H], and  the conversion of $^{22}$Ne to
Na with subsequent mixing of Na to the surface are independent of
metallicity,  HD\,179821 with [Fe/H] $\simeq +0.4$ and [Na/H]
also $\simeq +0.4$ for an unmixed star is
expected to have [Na/Fe] $\sim 0.7$, a value close to the
observed value in Table 6.

An alternative identification of HD\,179821 as a post-AGB star might
appear to be excluded on the grounds that key signatures of post-AGB stars
are absent, i.e., a C-rich and $s$-process atmosphere and envelope are
clearly not a feature of HD 179821. This exclusion supposes that all
AGB stars experience third dredge-up (i.e., envelope enrichment with
C and $s$-process nuclides) before evolving off the AGB. The case of
RV Tauri stars as post-AGB stars provides common examples of without
carbon and $s$-process enrichment. 

In such situations, when chemical
composition is not a definitive way to distinguish massive evolved
from post-AGB stars, other observed characteristics may be invoked. For example, the circumstellar CO
expansion velocity for HD\,179821 exceeds the typical velocity for post-AGB stars and likely requires a
very luminous (i.e., a massive) star in order to drive the expansion by radiation
pressure (Jura et al. 2001; Oudmaijer et al. 2008). But, obviously, among the other
characteristics the absolute luminosity plays a key
role, one looks forward to a precise trigonometrical parallax. Until then, HD\,179821 may be   
considered to be a warm Galactic supergiant like $\rho$ Cas and HR 8752.

% removed from the text after the reviewer's comment dated 27 May 2016 -- Timur
%One
%anticipates obtaining an accurate parallax before HD\,179821 explodes as
%a supernova or begins life as a white dwarf.

\section*{Acknowledgments}

DLL thanks to the Robert A. Welch Foundation of Houston, Texas for support through grant
F-634. VGK acknowledges the financial support by the Russian Foundation for Basic
Research (project 14-02-00291a). TS thanks TUBITAK--BIDEP for financial support
through its International Postdoctoral Research Scholarship Programme (1059B191500382 
). Christopher A. Sneden,
Paul Barklem, Michel Breger, Earle Luck, Monika Adamow, Eswar B. Reddy, and Sel\c{c}uk Bilir are
acknowledged for helpful remarks on the manuscript and useful correspondence.

\appendix
\section[]{Equivalent width measurements of neutral and ionized Fe lines over the 18 epochs listed in Table
1.}

\begin{table*}
\begin{center}
\caption{Equivalent width measurements of neutral and ionized Fe lines over the 18 epochs listed in Table
1.}
$$
\begin{array}{l|c|c|c|@{}r|@{}r|@{}r|@{}r|@{}r|@{}r|@{}r|@{}r|@{}r}
\hline
  &&   &  &$EW$ & $EW$ & $EW$&$EW$ &$EW$ & $EW$ &$EW$ &$EW$ &$EW$  \\
\cline{5-13}

$Wave.$  &$Spec.$&  $LEP$  &$loggf$   &\,$15Oct97$&\,$13Sep00$&\,$21Apr08$&\,$13Jun08$&\,$11Jul08$&\,$10Aug08$&\,$17Aug08$&\,$17Sep08$ &\,$14Apr09$ \\
\cline{1-1}
\cline{3-3}
\cline{5-13}
$(\AA)$  &     &  $(eV)$ &  &$(m$\AA$)$& $(m$\AA$)$&$(m$\AA$)$&$(m$\AA$)$&$(m$\AA$)$&$(m$\AA$)$&$(m$\AA$)$&$(m$\AA$)$&$(m$\AA$)$ \\

\hline \hline
4484.23&26.0 &  3.60 &  -0.86 & ...    &89.7 &  89.4   &118.6  &120    &128.8  &...    &...    &...	 \\
4602.95&26.0 &  1.49 &  -2.22 & 277.70 &142.9&  152.3  &156.3  &152.7  &136.5  &119.6  &...    &...	 \\
4643.47&26.0 &  3.65 &  -1.15 & 46.30  &28.5 &  37.8   &47.1   &32.5   &51.1   &29.9   &...    &...	 \\
5090.77&26.0 &  4.26 &  -0.44 & 76.70  &73.8 &  54.3   &105.1  &98.9   &118.4  &87.9   &...    &...	 \\
5353.37&26.0 &  4.10 &  -0.68 & 103.90 &55.6 &  64.0   &77.9   &...    &70.4   &66.2   &45.6   &63.6	 \\
5364.88&26.0 &  4.45 &  0.23  & 209.00 &148.8&  145.0  &167.4  &158.0  &152.4  &147.4  &117.3  &134.2	 \\
5367.48&26.0 &  4.42 &  0.44  & 215.40 &144.6&  196.0  &172.5  &161.0  &154.7  &149.8  &137.3  &160.5	 \\
5373.71&26.0 &  4.47 &  -0.84 & ...    &14.8 &  12.4   &17.4   &21.6   &19.3   &21.9   &11.4   &24.0	 \\
5393.18&26.0 &  3.24 &  -0.72 & ...    &103.7&  147.7  &130.9  &128.9  &121.8  &114.5  &88.4   &180.9	 \\
5434.53&26.0 &  1.01 &  -2.12 & 273.50 &213.4&  284.9  &257.9  &213.0  &228.3  &213.5  &190.3  &226.2	 \\
5569.63&26.0 &  3.42 &  -0.49 & 175.20 &108.9&  135.0  &152.9  &143.2  &138.5  &137.5  &107.5  &119.8	 \\
5572.85&26.0 &  3.40 &  -0.28 & 178.00 &154.4&  194.8  &201.9  &187.7  &178.7  &181.7  &162.1  &179.6	 \\
5816.38&26.0 &  4.55 &  -0.60 & 33.30  &36.0 &  38.0   &33.6   &30.4   &26.3   &35.7   &21.2   &26.8	 \\
6020.17&26.0 &  4.61 &  -0.21 & 57.30  &28.9 &  42.5   &48.7   &54.6   &41.1   &...    &27.8   &39.0	 \\
6024.07&26.0 &  4.55 &  -0.06 & 101.10 &68.3 &  87.3   &94.8   &92.2   &82.2   &...    &68.2   &77.6	 \\
6027.06&26.0 &  4.08 &  -1.09 & 51.70  &25.6 &  42.9   &46.3   &43.6   &34.8   &...    &22.8   &32.2	 \\
6065.49&26.0 &  2.61 &  -1.53 & ...    &52.9 &  64.6   &80.2   &63.6   &60.2   &...    &34.0   &50.5	 \\
6393.61&26.0 &  2.43 &  -1.58 & ...    &...  &  91.0   &99.1   &82.6   &76.3   &...    &47.0   &60.4	 \\
6411.66&26.0 &  3.65 &  -0.72 & ...    &70.1 &  88.1   &101.1  &91.8   &86.1   &...    &68.9   &76.9	 \\
6592.91&26.0 &  2.73 &  -1.47 & ...    &...  &  70.5   &97.2   &57.2   &60.9   &...    &...    &47.8	 \\
6841.34&26.0 &  4.61 &  -0.60 & ...    &...  &  37.8   &42.6   &39.5   &36.8   &...    &...    &...	 \\
4893.82&26.1 &  2.83 &  -4.27 & 272.30 &222.9&  228.5  &229.8  &212.7  &211.9  &221.8  &...    &...	 \\
5427.80&26.1 &  6.72 &  -1.58 & 74.10  &94.6 &  115.5  &86.9   &74.0   &72.0   &89.1   &83.0   &76.6	 \\
5813.67&26.1 &  5.57 &  -2.75 & 68.40  &65.6 &  83.2   &83.9   &78.4   &79.2   &85.5   &85.0   &77.6	 \\
5823.18&26.1 &  5.57 &  -2.99 & 59.10  &33.0 &  47.1   &66.8   &48.7   &45.7   &49.9   &37.9   &54.3	 \\
5991.38&26.1 &  3.15 &  -3.65 & 242.60 &...  &  215.8  &295    &245.2  &250.6  &239.1  &248.1  &253.5	 \\
6113.33&26.1 &  3.22 &  -4.23 & 133.70 &125.8&  135.1  &145.7  &130.5  &131.6  &...    &129.4  &138.9	 \\
6129.73&26.1 &  3.20 &  -4.74 & 118.80 &103.1&  90.1   &104.1  &96.4   &87.7   &...    &72.4   &67.3	 \\
6179.40&26.1 &  5.57 &  -2.80 & ...    &85.4 &  76.8   &82.4   &58.0   &69.1   &...    &56.8   &66.6	 \\
6446.40&26.1 &  6.22 &  -2.08 & ...    &...  &  86.0   &97.5   &98.7   &104.6  &...    &74.9   &92.9	 \\
\hline \hline
\end{array}
$$
\end{center} 
\end{table*}

\begin{table*}
\begin{center}
\caption{Equivalent width measurements of neutral and ionized Fe lines over the 18 epochs listed in Table
1.}
$$
\begin{array}{l|c|c|c|@{}r|@{}r|@{}r|@{}r|@{}r|@{}r|@{}r|@{}r|@{}r|@{}r}
\hline
  &&   &  &$EW$ & $EW$ & $EW$&$EW$ &$EW$ & $EW$ &$EW$ &$EW$ &$EW$ &$EW$ \\
\cline{5-14}
$Wave.$  &$Spec.$&  $LEP$  &$loggf$  
&\,$9May09$&\,$7Nov09$&\,$21Nov09$&\,$22May10$&\,$3Jun10$&\,$30Jul10$&\,$24Sep10$&\,$17Nov10$&\,$16May11$&\,$27Aug13$ \\
\cline{1-1}
\cline{3-3}
\cline{5-14}
(\AA)  &     &  (eV) &  &(m$\AA$)& (m$\AA$)& (m$\AA$)&(m$\AA$)&(m$\AA$)&(m$\AA$)&(m$\AA$)&(m$\AA$)&(m$\AA$)&(m$\AA$) \\

\hline \hline
4484.23&26.0 &  3.60 &  -0.86&102.8  &...    &174.2   &167.1  &...   & 74.8  &...    &...    &151.2&182.9   \\
4602.95&26.0 &  1.49 &  -2.22&151.2  &...    &146.6   &197.6  &...   &115.1  &...    &...    &155.6&188.6   \\
4643.47&26.0 &  3.65 &  -1.15& 47.8  &...    & 34.3   & 45.8  &...   & 20.4  &...    &...    & 36.3&55.0   \\
5090.77&26.0 &  4.26 &  -0.44&101.1  &...    &115.0   &121.2  &...   & 60.9  &...    &...    & 69.5&110.3   \\
5353.37&26.0 &  4.10 &  -0.68& 77.2  &  60.7 & 69.0   &102.6  & 93.3 & 56.0  & 70.4  & 94.7  & 72.4&97.7   \\
5364.88&26.0 &  4.45 &   0.23&157.0  & 147.5 &179.8   &192.0  &181.5 &121.1  &148.0  &218.8  &178.7&203.0   \\
5367.48&26.0 &  4.42 &   0.44&190.7  & 146.4 &160.3   &188.1  &188.0 &140.6  &158.9  &221.0  &168.1&194.0   \\
5373.71&26.0 &  4.47 &  -0.84& 26.5  &  15.9 & 27.4   &...    & 26.6 & 24.7  & 19.7  & 22.5  & 27.1&36.7   \\
5393.18&26.0 &  3.24 &  -0.72&133.9  &  98.9 &126.6   &168.7  &153.1 &102.5  &112.0  &157.0  &127.2&159.4   \\
5434.53&26.0 &  1.01 &  -2.12&234.4  & 218.0 &196.6   &316.9  &315.2 &200.1  &238.8  &301.7  &230.0&285.6   \\
5569.63&26.0 &  3.42 &  -0.49&149.4  & 126.5 &143.4   &154.0  &175.7 &...    &137.5  &176.6  &148.5&187.9   \\
5572.85&26.0 &  3.40 &  -0.28&209.4  & 194.2 &185.6   &227.1  &219.5 &...    &181.9  &224.0  &199.6&230.5   \\
5816.38&26.0 &  4.55 &  -0.60& 35.0  &...    & 23.8   & 45.0  & 34.8 &...    & 27.1  & 39.1  & 28.4&37.9   \\
6020.17&26.0 &  4.61 &  -0.21& 56.9  &...    & 46.5   & 66.4  & 62.1 &...    & 43.7  & 75.1  & 46.2&87.9   \\
6024.07&26.0 &  4.55 &  -0.06& 96.8  &...    & 96.4   &107.6  &102.8 &...    & 87.5  &108.3  & 85.1&118.6   \\
6027.06&26.0 &  4.08 &  -1.09& 43.5  &...    & 33.3   & 59.5  & 52.0 &...    & 33.3  & 52.8  & 39.1&55.8   \\
6065.49&26.0 &  2.61 &  -1.53& 70.0  &...    & 65.1   & 93.4  & 93.5 &...    & 62.1  & 87.9  & 60.9&91.9   \\
6393.61&26.0 &  2.43 &  -1.58& 87.3  &...    & 83.3   &124.7  &112.4 &...    & 70.9  &119.2  & 68.8&94.6   \\
6411.66&26.0 &  3.65 &  -0.72& 98.6  &  66.6 & 88.9   &111.8  &100.3 &...    & 86.6  &104.8  & 87.4&130.3   \\
6592.91&26.0 &  2.73 &  -1.47& 71.7  &...    & 58.7   &106.9  & 87.7 &...    & 64.6  &102.0  & 68.1&...   \\
6841.34&26.0 &  4.61 &  -0.60& 50.3  &...    & 18.8   & 40.1  &...   &...    &...    &...    & 51.3&...   \\
4893.82&26.1 &  2.83 &  -4.27&238.5  &...    &228.4   &212.5  &...   &230.1  &...    &...    &252.2&266.1   \\
5427.80&26.1 &  6.72 &  -1.58& 91.1  &  80.9 & 95.4   & 91.0  & 85.4 & 98.1  & 81.1  & 95.7  & 88.3&101.7   \\
5813.67&26.1 &  5.57 &  -2.75& 86.4  &...    & 83.7   & 93.1  & 70.9 &...    & 88.3  & 82.8  & 94.9&87.1   \\
5823.18&26.1 &  5.57 &  -2.99& 67.9  &...    & 44.4   & 63.2  & 50.8 &...    & 38.5  & 46.0  & 42.3&49.2   \\
5991.38&26.1 &  3.15 &  -3.65&223.0  &...    &266.0   &281.5  &286.0 &...    &240.7  &264.4  &268.7&...   \\
6113.33&26.1 &  3.22 &  -4.23&137.7  &...    &145.4   &147.2  &152.2 &...    &133.5  &143.4  &132.6&...   \\
6129.73&26.1 &  3.20 &  -4.74&111.0  &...    &105.6   &118.2  &      &...    & 89.8  &101.8  & 95.5&106.5   \\
6179.40&26.1 &  5.57 &  -2.80& 69.4  &...    & 76.3   & 68.3  & 74.7 &...    & 71.2  & 80.6  & 80.2&92.7   \\
6446.40&26.1 &  6.22 &  -2.08& 89.4  &  89.6 & 97.1   &103.2  & 91.5 &...    & 85.4  & 86.6  & 74.3&...   \\
\hline \hline
\end{array}
$$
\end{center} 
\end{table*}

\label{lastpage}
\end{document}